\documentclass[aps,prd,epsfig,preprint,amsmath,amssymb,superscriptaddress,showpacs,nofootinbib]{revtex4}
\usepackage{multirow}
\usepackage{graphicx,bm}
\usepackage[usenames]{color}
\usepackage{float}
\usepackage{caption}
\usepackage{subcaption}
\usepackage{slashed}
\begin{document}
\draft
\title{Probe of the anomalous neutral triple gauge couplings in photon-induced collision at future muon colliders}

\author{S. Spor}
\email[]{serdar.spor@beun.edu.tr}
\affiliation{Department of Medical Imaging Techniques, Zonguldak B\"{u}lent Ecevit University, 67100, Zonguldak, Turkey.}

\date{\today}

\begin{abstract}

The anomalous $ZZ\gamma$ and $Z\gamma\gamma$ neutral triple gauge couplings occurred by dimension-eight operators are investigated through the process $\mu^+\mu^-\,\rightarrow\,\mu^+\gamma^*\mu^-\,\rightarrow\,\mu^+ Z(\nu\bar{\nu})\mu^-$ at the muon collider with $\sqrt{s}=3$, $6$, $10$ and $14$ TeV. The charged lepton pseudo-rapidity, the charged lepton transverse momentum and the transverse missing energy distributions are taken in consideration for the final state of the process in the analysis. The sensitivities of the anomalous couplings are obtained at $95\%$ Confidence Level with integrated luminosity of ${\cal L}_{\text{int}}=1$, $4$, $10$ and $20$ ab$^{-1}$, respectively, according to center-of-mass energies of muon collider taking into account the effects of systematic uncertainties $0\%$, $3\%$ and $5\%$. The best limits of anomalous $C_{BB}/{\Lambda^4}$, $C_{BW}/{\Lambda^4}$, $C_{\widetilde{B}W}/{\Lambda^4}$ and $C_{WW}/{\Lambda^4}$ couplings without systematic uncertainty at center-of-mass energy of 14 TeV and integrated luminosity of 20 ab$^{-1}$ are found to be [-0.01026; 0.00636] TeV$^{-4}$, [-0.02482; 0.03053] TeV$^{-4}$, [-0.01830; 0.02510] TeV$^{-4}$, [-0.06981; 0.07387] TeV$^{-4}$, respectively.

\end{abstract}


\maketitle

\section{Introduction}

Nowadays, some of the shortcomings that the Standard Model (SM) still cannot answer reveal the existence of new physics beyond the SM, and the necessity of a high-energy collider to search for evidence has become clear. After the discovery of the Higgs boson at the LHC, the question of which collider will be built in the future has come to the fore. Continuing particle physics research in the post-LHC period depends on analyzing new future machine options. In this study, the physics potential and the theoretical advantages of future multi-TeV muon colliders are emphasized.

There are many remarkable features of the multi-TeV muon collider. The fact that muons are heavier than electrons causes less synchrotron radiation to be produced, making it easier to accelerate them to high energies with a circular collider. Due to the nature of lepton colliders, muons use up all of their center-of-mass energy since they are not compound particles like protons. Lepton colliders also have much smaller backgrounds and cleaner environment than hadron colliders. In addition to these theoretical advantages, experimental studies such as the Muon Accelerator Program (MAP) \cite{Palmer:2014asd,Delahaye:2013hvz}, the Muon Ionization Cooling Experiment (MICE) \cite{Bogomilov:2020twm}, and the Low Emittance Muon Accelerator (LEMMA) \cite{Antonelli:2016ezx} have greatly increased the motivation to build a future muon collider. In the light of experimental predictions for muon colliders, interest in phenomenological studies has also increased day by day. These studies are generally examined under the headings such as the electroweak parton distribution function (EW PDF) formalism in muon collisions \cite{Han:2021gbv,Han:2022rws,Ruiz:2021oec}, measuring the Higgs boson couplings \cite{Costantini:2020tkp,Chiesa:2020yhn,Han:2021pas,Chen:2021pln,Chen:2022ygc}, BSM studies with new scalars \cite{Eichten:2014evo,Chakrabarty:2015gwe,Buttazzo:2018wmg,Bandyopadhyay:2021lja,Han:2021hrq,Liu:2021gtr}, minimal dark matter \cite{Han:2021twq,Capdevilla:2021xku,Bottaro:2021res,Han:2022epx}, the muon $g-2$ anomaly \cite{Yin:2020gre,Capdevilla:2021ooc,Buttazzo:2021ooc,Dermisek:2021sdf,Capdevilla:2022yza}, the neutral current B-meson anomalies \cite{Huang:2021edc,Asadi:2021wsd} and investigation of anomalous neutral triple gauge couplings (aNTGC) \cite{Spor:2022rcx,Senol:2022evm} and anomalous quartic gauge couplings (aQGC) \cite{Yang:2022ubn,Yang:2022emz}.

Four different scenarios of muon collider are considered in this paper. The center-of-mass energies for these scenarios, along with the energies of incoming muon and the integrated luminosities, are given in Table~\ref{tab1} and these values have been used in previous studies for the muon collider \cite{Han:2021twq,Ali:2021wsd,Chiesa:2021tyr}.

\begin{table}[H]
\centering
\caption{Energy and integrated luminosity in scenarios of muon collider.}
\label{tab1}
\begin{tabular}{p{3cm}p{1.5cm}p{1.5cm}p{1.5cm}p{1.5cm}}
\hline \hline
$\sqrt{s}$ (TeV) & 3 & 6 & 10 & 14\\ \hline
$E_{\mu}$ (TeV) & 1.5 & 3 & 5 & 7\\ \hline
${\cal L}_{\text{int}}$ (ab$^{-1}$) & 1 & 4 & 10 & 20\\  \hline \hline
\end{tabular}
\end{table}

Since the Z boson has no electric charge, there is no coupling between the Z boson and the photon at the tree level in the SM. Therefore, the presence of aNTGCs between the photon and the $Z$ boson ($Z\gamma\gamma$ and $ZZ\gamma$) is a clue to detect deviation from the SM prediction and plays a key role in the exploration of new physics beyond the SM \cite{Spor:2022omz}. This possible deviations on aNTGC interactions are investigated with a model-independent framework in the Effective Field Theory (EFT). The effective Lagrangian of NTGC consists of the SM Lagrangian with dimension-four operators and new physics contributions beyond the SM with dimension-eight operators and can be written as \cite{Degrande:2014ydn}
 
\begin{eqnarray}
\label{eq.1} 
{\cal L}^{\text{nTGC}}={\cal L}_{\text{SM}}+\sum_{i}\frac{C_i}{\Lambda^{4}}({\cal O}_i+{\cal O}_i^\dagger)
\end{eqnarray}

{\raggedright where new physics terms are suppressed by the inverse powers of the new physics scale $\Lambda$ and ${\cal O}_i$ defines the four operators given below}

\begin{eqnarray}
\label{eq.2} 
{\cal O}_{\widetilde{B}W}=iH^{\dagger} \widetilde{B}_{\mu\nu}W^{\mu\rho} \{D_\rho,D^\nu \}H,
\end{eqnarray}
\begin{eqnarray}
\label{eq.3} 
{\cal O}_{BW}=iH^\dagger B_{\mu\nu}W^{\mu\rho} \{D_\rho,D^\nu \}H,
\end{eqnarray}
\begin{eqnarray}
\label{eq.4} 
{\cal O}_{WW}=iH^\dagger W_{\mu\nu}W^{\mu\rho} \{D_\rho,D^\nu \}H,
\end{eqnarray}
\begin{eqnarray}
\label{eq.5} 
{\cal O}_{BB}=iH^\dagger B_{\mu\nu}B^{\mu\rho} \{D_\rho,D^\nu \}H.
\end{eqnarray}

The first of these dimension-eight operators is CP-even and the last three are CP-odd. $B_{\mu\nu}$ and $W^{\mu\nu}$ are the field strength tensors, $\widetilde{B}_{\mu\nu}$ is the dual $B$ strength tensor, $H$ is the Higgs field and $D_\mu$ is the covariant derivative. The following expressions are written for the definition of operators:

\begin{eqnarray}
\label{eq.6} 
B_{\mu\nu}=\left(\partial_\mu B_\nu - \partial_\nu B_\mu\right),
\end{eqnarray}
\begin{eqnarray}
\label{eq.7} 
W_{\mu\nu}=\sigma^i\left(\partial_\mu W_\nu^i - \partial_\nu W_\mu^i + g\epsilon_{ijk}W_\mu^j W_\nu^k\right),
\end{eqnarray}
\begin{eqnarray}
\label{eq.8} 
D_\mu \equiv \partial_\mu - i\frac{g^\prime}{2}B_\mu Y - ig_W W_\mu^i\sigma^i.
\end{eqnarray}

Although the dimension-six operators do not contribute new physics on aNTGC at tree-level, their contribution at one-loop is of the order ${\alpha \hat{s}}/{4\pi\Lambda^2}$. The tree-level contribution of the dimension-eight operators is of the order ${\upsilon^2\hat{s}}/{\Lambda^4}$. In conclusion, the contribution of the dimension-eight operators is more dominant than the contribution of the dimension-six operator at one-loop with $\Lambda \lesssim \sqrt{4\pi\hat{s}/\alpha}$ \cite{Degrande:2014ydn}.

Effective Lagrangian for aNTGC with dimension-six and dimension-eight operators is written by \cite{Gounaris:2000svs}

\begin{eqnarray}
\label{eq.9} 
\begin{split}
{\cal L}_{\text{aNTGC}}^{\text{dim-6,8}}=&\frac{g_e}{m_Z^2}\Bigg[-[f_4^\gamma(\partial_\mu F^{\mu\beta})+f_4^Z(\partial_\mu Z^{\mu\beta})]Z_\alpha (\partial^\alpha Z_\beta)+[f_5^\gamma(\partial^\sigma F_{\sigma\mu})+f_5^Z (\partial^\sigma Z_{\sigma\mu})]\widetilde{Z}^{\mu\beta}Z_\beta  \\
&-[h_1^\gamma (\partial^\sigma F_{\sigma\mu})+h_1^Z (\partial^\sigma Z_{\sigma\mu})]Z_\beta F^{\mu\beta}-[h_3^\gamma(\partial_\sigma F^{\sigma\rho})+h_3^Z(\partial_\sigma Z^{\sigma\rho})]Z^\alpha \widetilde{F}_{\rho\alpha}   \\
&-\bigg\{\frac{h_2^\gamma}{m_Z^2}[\partial_\alpha \partial_\beta \partial^\rho F_{\rho\mu}]+\frac{h_2^Z}{m_Z^2}[\partial_\alpha \partial_\beta(\square+m_Z^2)Z_\mu]\bigg\}Z^\alpha F^{\mu\beta}   \\
&+\bigg\{\frac{h_4^\gamma}{2m_Z^2}[\square\partial^\sigma F^{\rho\alpha}]+\frac{h_4^Z}{2m_Z^2}[(\square+m_Z^2)\partial^\sigma Z^{\rho\alpha}]\bigg\}Z_\sigma\widetilde{F}_{\rho\alpha}\Bigg],
\end{split}
\end{eqnarray}

{\raggedright where $\widetilde{Z}_{\mu\nu}=1/2\epsilon_{\mu\nu\rho\sigma}Z^{\rho\sigma}$ $(\epsilon^{0123}=+1)$ with field strength tensor $Z_{\mu\nu}=\partial_\mu Z_\nu - \partial_\nu Z_\mu$ and similarly for the electromagnetic field tensor $F_{\mu\nu}$. Three CP-violating couplings are $f_4^V$, $h_1^V$, $h_2^V$ while three CP-conserving couplings are $f_5^V$, $h_3^V$, $h_4^V$ with $(V=\gamma$, $Z)$ \cite{Gounaris:2000thb}. All couplings are zero in the SM at tree-level. At the Lagrangian in Eq.~(\ref{eq.9}), the couplings $h_2^V$ and $h_4^V$ correspond to dimension-eight and the other four couplings to dimension-six \cite{Moyotl:2015qsf}.} 

The couplings of the effective Lagrangian in Eq.~(\ref{eq.9}) are associated to the couplings of the operators in Eqs.~(\ref{eq.2}-\ref{eq.5}) on the gauge invariance under the $SU(2)_L \times U(1)_Y$ group \cite{Rahaman:2020fdf}. The CP-conserving anomalous couplings with two on-shell $Z$ bosons and one off-shell $V=\gamma$ or $Z$ boson for the $ZZV$ coupling are written by \cite{Degrande:2014ydn}

\begin{eqnarray}
\label{eq.10} 
f_5^Z=0,
\end{eqnarray}
\begin{eqnarray}
\label{eq.11} 
f_5^\gamma=\frac{\upsilon^2 m_Z^2}{4c_\omega s_\omega} \frac{C_{\widetilde{B}W}}{\Lambda^4}
\end{eqnarray}

{\raggedright and the CP-violating anomalous couplings by}

\begin{eqnarray}
\label{eq.12} 
f_4^Z=\frac{m_Z^2 \upsilon^2 \left(c_\omega^2 \frac{C_{WW}}{\Lambda^4}+2c_\omega s_\omega \frac{C_{BW}}{\Lambda^4}+4s_\omega^2 \frac{C_{BB}}{\Lambda^4}\right)}{2c_\omega s_\omega},
\end{eqnarray}
\begin{eqnarray}
\label{eq.13} 
f_4^\gamma=-\frac{m_Z^2 \upsilon^2 \left(-c_\omega s_\omega \frac{C_{WW}}{\Lambda^4}+\frac{C_{BW}}{\Lambda^4}(c_\omega^2-s_\omega^2)+4c_\omega s_\omega \frac{C_{BB}}{\Lambda^4}\right)}{4c_\omega s_\omega}.
\end{eqnarray}

The CP-conserving anomalous couplings with one on-shell $Z$ boson, one on-shell photon and one off-shell $V=\gamma$ or $Z$ boson for the $Z\gamma V$ coupling are written by \cite{Degrande:2014ydn}

\begin{eqnarray}
\label{eq.14} 
h_3^Z=\frac{\upsilon^2 m_Z^2}{4c_\omega s_\omega} \frac{C_{\widetilde{B}W}}{\Lambda^4},
\end{eqnarray}
\begin{eqnarray}
\label{eq.15} 
h_4^Z=h_3^\gamma=h_4^\gamma=0 ,
\end{eqnarray}

{\raggedright and the CP-violating anomalous couplings by}

\begin{eqnarray}
\label{eq.16} 
h_1^Z=\frac{m_Z^2 \upsilon^2 \left(-c_\omega s_\omega \frac{C_{WW}}{\Lambda^4}+\frac{C_{BW}}{\Lambda^4}(c_\omega^2-s_\omega^2)+4c_\omega s_\omega \frac{C_{BB}}{\Lambda^4}\right)}{4c_\omega s_\omega},
\end{eqnarray}
\begin{eqnarray}
\label{eq.17} 
h_2^Z=h_2^\gamma=0,
\end{eqnarray}
\begin{eqnarray}
\label{eq.18} 
h_1^\gamma=-\frac{m_Z^2 \upsilon^2 \left(s_\omega^2 \frac{C_{WW}}{\Lambda^4}-2c_\omega s_\omega \frac{C_{BW}}{\Lambda^4}+4c_\omega^2 \frac{C_{BB}}{\Lambda^4}\right)}{4c_\omega s_\omega}.
\end{eqnarray}

The anomalous couplings with two on-shell photons and one off-shell $Z$ boson are not considered because it is prohibited by the Landau-Yang theorem \cite{Landau:1948gbe,Yang:1950hyb}. In Eqs.~(\ref{eq.11}-\ref{eq.14},\ref{eq.16},\ref{eq.18}) the coefficients $C_{BB}/{\Lambda^4}$, $C_{BW}/{\Lambda^4}$, $C_{\widetilde{B}W}/{\Lambda^4}$, $C_{WW}/{\Lambda^4}$ are dimension-eight aNTGC and $C_{\widetilde{B}W}/{\Lambda^4}$ is CP-conserving coupling while $C_{BB}/{\Lambda^4}$, $C_{BW}/{\Lambda^4}$, $C_{WW}/{\Lambda^4}$ are CP-violating couplings.

In this study, the sensitivity of dimension-eight $C_{BB}/{\Lambda^4}$, $C_{BW}/{\Lambda^4}$, $C_{\widetilde{B}W}/{\Lambda^4}$ and $C_{WW}/{\Lambda^4}$ couplings in anomalous $ZZ\gamma$ and $Z\gamma\gamma$ vertices is investigated at multi-TeV muon colliders with the process $\mu^+\mu^-\,\rightarrow\,\mu^+\gamma^*\mu^-\,\rightarrow\,\mu^+ Z\mu^-$. Since no dimension-six operator induces aNTGC at this process, the new physics contribution to these interactions comes from the dimension-eight operators, and the effects of the new physics depend on the presence of these operators and the extent of their contributions. There are many new physics studies on the dimension-eight operators describing aNTGC \cite{Senol:2018gvg,Senol:2019ybv,Ellis:2020ekm,Yilmaz:2020ser,Senol:2020hbh,Ellis:2021rop,Fu:2021jec,Yilmaz:2021dbm,Spor:2022omz,Senol:2022evm}. The current experimental limits on dimension-eight aNTGCs at the $95\%$ C.L. are studied through process $pp\rightarrow Z\gamma$ and $pp\rightarrow ZZ$ with center-of-mass energy of 13 TeV at the CERN and given in Table \ref{tab2}.

\begin{table}[H]
\caption{The current experimental limits on dimension-eight aNTGCs at the $95\%$ C.L..}
\label{tab2}
\begin{ruledtabular}
\begin{tabular}{lcccc}
\multirow{2}{*}{Experimental limits} & \multicolumn{4}{c}{Couplings (TeV$^{-4}$)}\\
 & ${C_{\widetilde{B}W}}/{\Lambda^4}$ & ${C_{WW}}/{\Lambda^4}$ & ${C_{BW}}/{\Lambda^4}$ & ${C_{BB}}/{\Lambda^4}$\\ \hline
ATLAS \cite{Aaboud:2018ybz} $Z\gamma\rightarrow\nu\bar{\nu}\gamma$ & \multirow{2}{*}{-1.1; 1.1} & \multirow{2}{*}{-2.3; 2.3} & \multirow{2}{*}{-0.65; 0.64} & \multirow{2}{*}{-0.24; 0.24}\\ 
($\sqrt{s}=13$ TeV, ${\cal L}_{\text{int}}=36.1$ fb$^{-1}$) & & & & \\ \hline
ATLAS \cite{Aaboud:2018onm} $ZZ\rightarrow\ell^+\ell^-\ell^{\prime+}\ell^{\prime-}$ & \multirow{2}{*}{-5.9; 5.9} & \multirow{2}{*}{-3.0; 3.0} & \multirow{2}{*}{-3.3; 3.3} & \multirow{2}{*}{-2.7; 2.8}\\ 
($\sqrt{s}=13$ TeV, ${\cal L}_{\text{int}}=36.1$ fb$^{-1}$) & & & & \\ \hline
CMS \cite{Sirunyan:2021edk} $ZZ\rightarrow\ell^+\ell^-\ell^{\prime+}\ell^{\prime-}$ & \multirow{2}{*}{-2.3; 2.5} & \multirow{2}{*}{-1.4; 1.2} & \multirow{2}{*}{-1.4; 1.3} & \multirow{2}{*}{-1.2; 1.2}\\ 
($\sqrt{s}=13$ TeV, ${\cal L}_{\text{int}}=137$ fb$^{-1}$) & & & & \\
\end{tabular}
\end{ruledtabular}
\end{table}

\section{Cross-sections measurements at the muon collider} \label{Sec2}

\begin{figure}[H]
\centering
\includegraphics[scale=0.7]{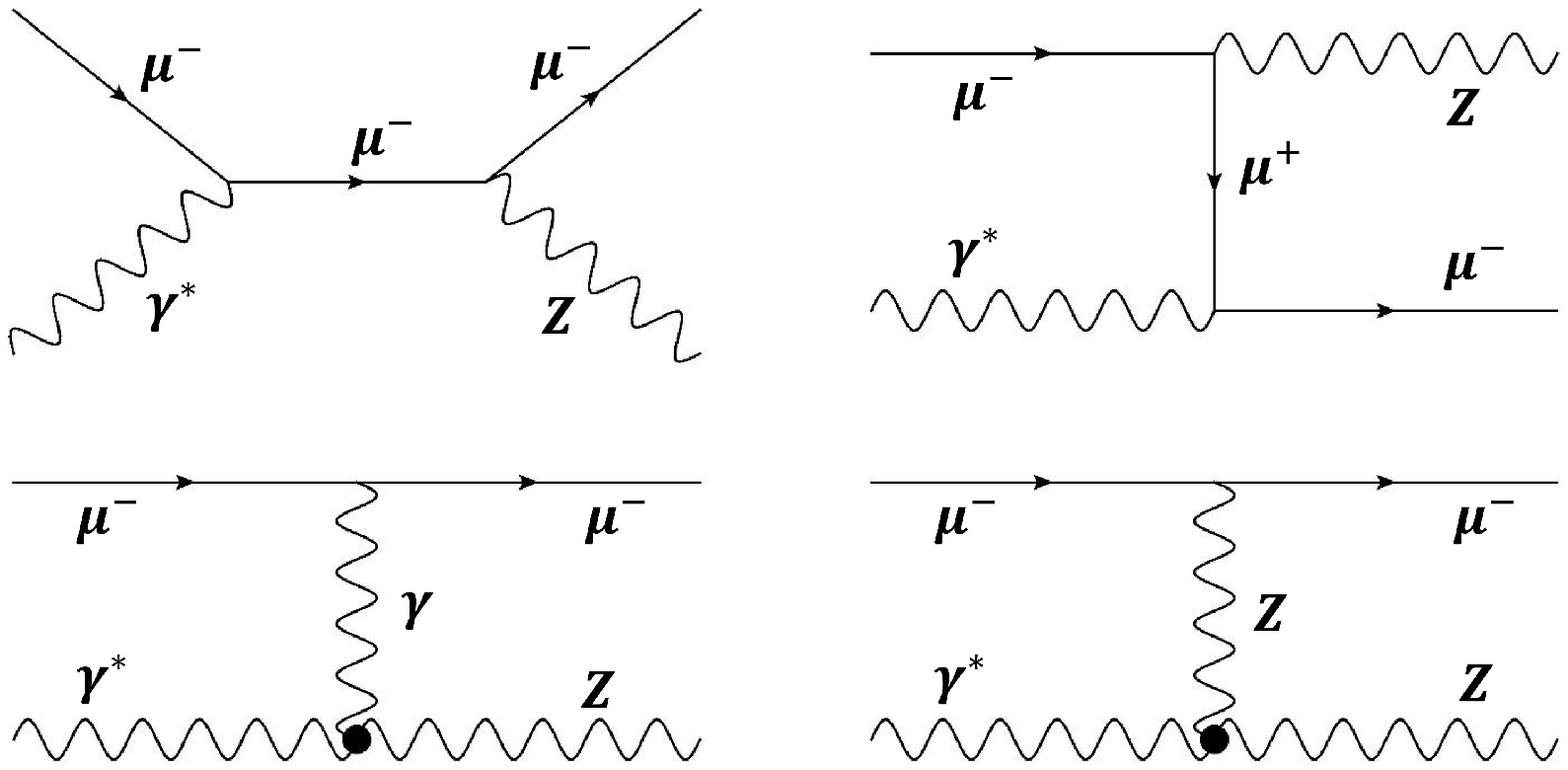}
\caption{The Feynman diagrams of the process $\mu^-\gamma^*\,\rightarrow\,Z\mu^-$. 
\label{fig1}}
\end{figure}

Four Feynman diagrams at the tree level for the process $\mu^-\gamma^*\,\rightarrow\,Z\mu^-$ are shown in Fig.~\ref{fig1}. The first two diagrams include SM contributions, while the others include new physics contributions beyond the SM with the anomalous $Z\gamma\gamma$ and $ZZ\gamma$ couplings. In this study, the decay of the $Z$ boson into neutrino-antineutrino pair is considered. Because, the process involving the decay of the $Z$ boson into neutrino-antineutrino pair is advantageous, since the hadron channel with a large multi-jet background does not have clean data, and the decay of the $Z$ boson into charged leptons has a low branching ratio compared to the neutrinos decay.

In this study, $\mu^-\gamma^*$ collisions occur when the photon $\gamma^*$ emitted from the $\mu^+$ beam collides with other muon beam. However, the fact that the photon $\gamma^*$ can also be emitted from the $\mu^-$ instead of the $\mu^+$ should definitely not be overlooked. In Equivalent Photon Approximation (EPA), the spectrum of photon emitted from muon is written by \cite{Budnev:1975kyp,Koksal:2019ybm}:

\begin{eqnarray}
\label{eq.19}
\begin{split}
f_{\gamma^{*}}(x)=&\, \frac{\alpha}{\pi E_{\mu}}\Bigg\{\left[\frac{1-x+x^{2}/2}{x}\right]\text{log}\left(\frac{Q_{\text{max}}^{2}}{Q_{\text{min}}^{2}}\right)-\frac{m_{\mu}^{2}x}{Q_{\text{min}}^{2}}\left(1-\frac{Q_{\text{min}}^{2}}{Q_{\text{max}}^{2}}\right)
\\
&-\frac{1}{x}\left[1-\frac{x}{2}\right]^{2}\text{log}\left(\frac{x^{2}E_{\mu}^{2}+Q_{\text{max}}^{2}}{x^{2}E_{\mu}^{2}+Q_{\text{min}}^{2}}\right)\Bigg\}\,,
\end{split}
\end{eqnarray}

{\raggedright where $x=E_{\gamma^{*}}/E_{\mu}$ and $Q_{\text{min}}^{2}=m_{\mu}^{2}x^{2}/(1-x)$. $Q_{\text{max}}^{2}$ is maximum virtuality of the photon.}

The total cross-section of the process $\mu^+\mu^-\,\rightarrow\,\mu^+\gamma^*\mu^-\,\rightarrow\,\mu^+ Z\mu^-$ is obtained by integrating over both the cross-section of the subprocess $\mu^-\gamma^*\,\rightarrow\,Z\mu^-$ and the spectrum of photon emitted from muon. The total cross-section is given as follows:

\begin{eqnarray}
\label{eq.20}
\sigma\left( \mu^+\mu^-\,\rightarrow\,\mu^+\gamma^*\mu^-\,\rightarrow\,\mu^+ Z\mu^- \right)=\int f_{\gamma^{*}}(x)\hat{\sigma}\left({\mu^-\gamma^*\,\rightarrow\,Z\mu^-}\right) dx\,.
\end{eqnarray}

Any non-zero anomalous couplings of $C_{BB}/{\Lambda^4}$, $C_{BW}/{\Lambda^4}$, $C_{\widetilde{B}W}/{\Lambda^4}$, $C_{WW}/{\Lambda^4}$ and SM contributions in the process $\mu^-\gamma^*\,\rightarrow\,Z(\nu\bar{\nu})\mu^-$ are considered a signal. In the analysis of the process $\mu^-\gamma^*\,\rightarrow\,Z\mu^-\,\rightarrow\,\nu\bar{\nu}\mu^-$, it is seen that the final state topology of the signal process consists of a isolated lepton and missing transverse energy. The SM background process ($\mu^-\gamma^*\,\rightarrow\,Z(\nu\bar{\nu})\mu^-$) has the same final state as the signal process, including only SM contributions from the first two Feynman diagrams in Fig.~\ref{fig1}. Other backgrounds: (1) The production of a $W$ boson that decays into a charged lepton and a neutrino is considered ($\mu^-\gamma^*\,\rightarrow\,W^-(\ell^- \bar{\nu})\nu_\mu$). Thus in the final state, it include one charged lepton ($\ell^- = \mu^-$) and two neutrinos. (2) When the outgoing photon is outside angular acceptance and therefore missing along the pipeline, the process $\mu^-\gamma^*\rightarrow \mu^-\gamma$ is expected to contribute to the background. (3) By performing $ZZ$ diboson production, a background that decays into two neutrinos and two charged leptons ($\ell^\pm = \mu^\pm$) in the final state is considered with $\mu^-\mu^+\,\rightarrow\,Z(\nu\bar{\nu})Z(\ell^- \ell^+)$. (4) By performing WW diboson production, a background with leptonic decay of both W bosons in the final state is considered with $\mu^-\mu^+\,\rightarrow\,W^-(\ell^- \bar{\nu})W^+(\ell^+ \nu)$, where $\ell^- = \mu^-$ and $\ell^+ = e^+,\mu^+$.

All signal and background events are generated using {\sc MadGraph5}$\_$aMC@NLO \cite{Alwall:2014cvc} with $500$k events for each. Kinematic cuts at the generator level are given in Table~\ref{tab3} and these cuts are called basic cuts: photon pseudo-rapidity $|\eta^{\gamma}|$, charged lepton pseudo-rapidity $|\eta^{\ell}|$, photon transverse momentum $p^\gamma_T$, charged lepton transverse momentum $p^\ell_T$, minimum distance between photons and leptons $\Delta R^{\gamma\ell}_{\text{min}}$. Also, as a result of the decay of one of the $Z$ bosons into a lepton-antilepton pair and the leptonic decay of both $W$ bosons in the $\mu^+\mu^-\rightarrow ZZ, WW$ processes, an additional lepton is present in both background processes. These additional leptons are vetoed with $p^\ell_T>10$ GeV (and $|\eta^\ell|<2.5$) to reduce the contribution from standard model processes that produce a higher number of leptons. It is necessary to apply more powerful kinematic cuts to distinguish the signal from the relevant backgrounds and after the suitable cuts selected, the relevant background is suppressed. Therefore, as suitable cuts, we can use the charged lepton pseudo-rapidity $\eta^{\ell}$, the charged lepton transverse momentum $p^\ell_T$ and the transverse missing energy $\slashed{E}_T$ in Table~\ref{tab3} for the final state of the process $\mu^-\gamma^*\,\rightarrow\,\nu\bar{\nu}\mu^-$. These cuts are called selected cuts. 

\begin{table}[H]
\centering
\caption{The values of the kinematic cuts applied in the analysis.}
\label{tab3}
\begin{tabular}{p{5cm}p{3cm}}
\hline \hline
Basic Cuts & Selected Cuts\\ \hline
$|\eta^{\gamma}|<2.5$ & $|\eta^{\ell}|<2.5$\\ 
$|\eta^{\ell}|<2.5$ & $p^\ell_T>30$ GeV\\ 
$p^\gamma_T>10$ GeV & $\slashed{E}_T>200$ GeV\\  
$p^\ell_T>10$ GeV & \\ 
$\Delta R^{\gamma\ell}_{\text{min}}>0.4$ & \\  \hline\hline
\end{tabular}
\end{table}

\begin{figure}[H]
\centering
\begin{subfigure}{0.5\linewidth}
\includegraphics[width=\linewidth]{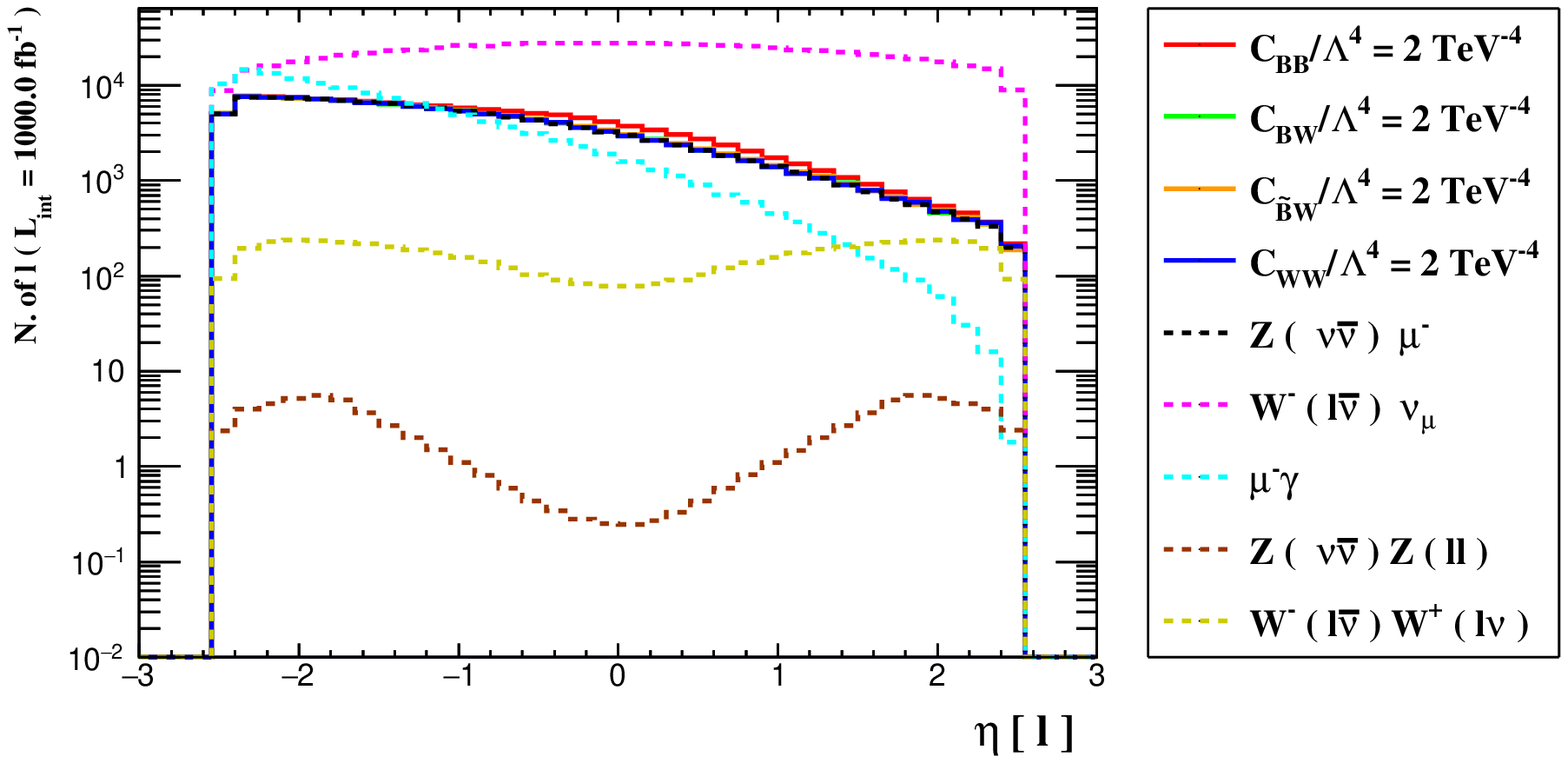}
\caption{}
\label{fig2:a}
\end{subfigure}\hfill
\begin{subfigure}{0.5\linewidth}
\includegraphics[width=\linewidth]{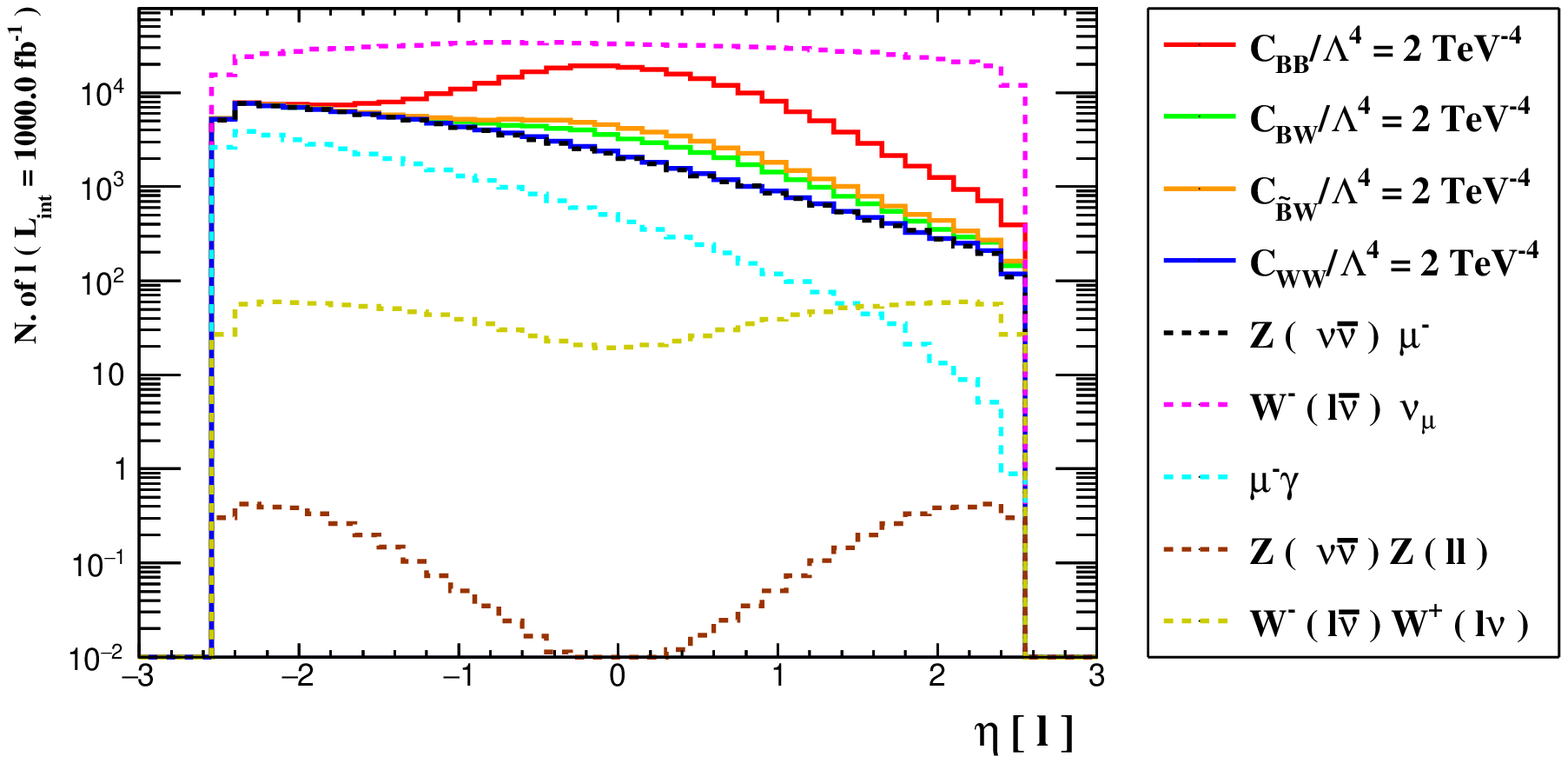}
\caption{}
\label{fig2:b}
\end{subfigure}\hfill

\begin{subfigure}{0.5\linewidth}
\includegraphics[width=\linewidth]{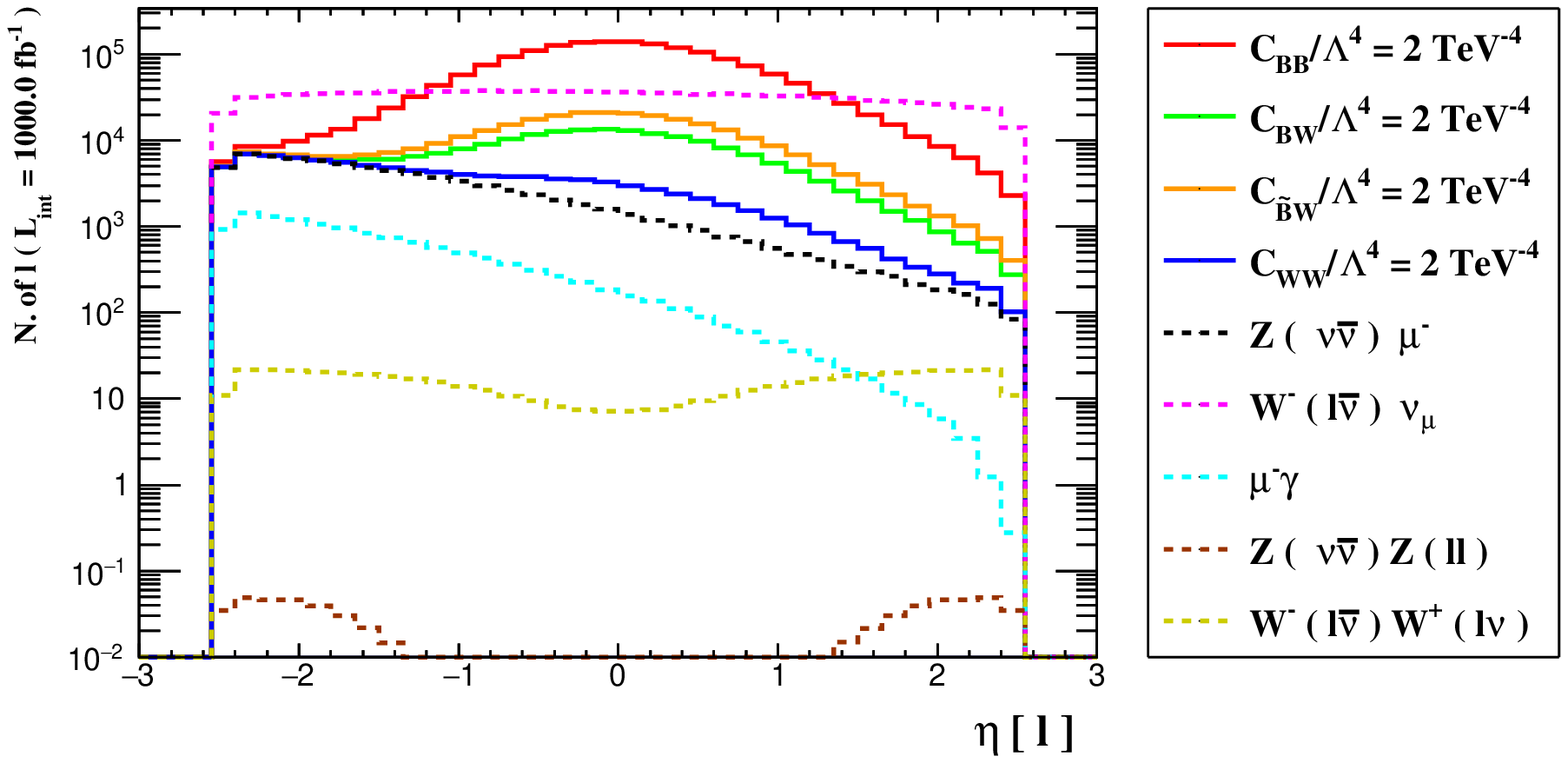}
\caption{}
\label{fig2:c}
\end{subfigure}\hfill
\begin{subfigure}{0.5\linewidth}
\includegraphics[width=\linewidth]{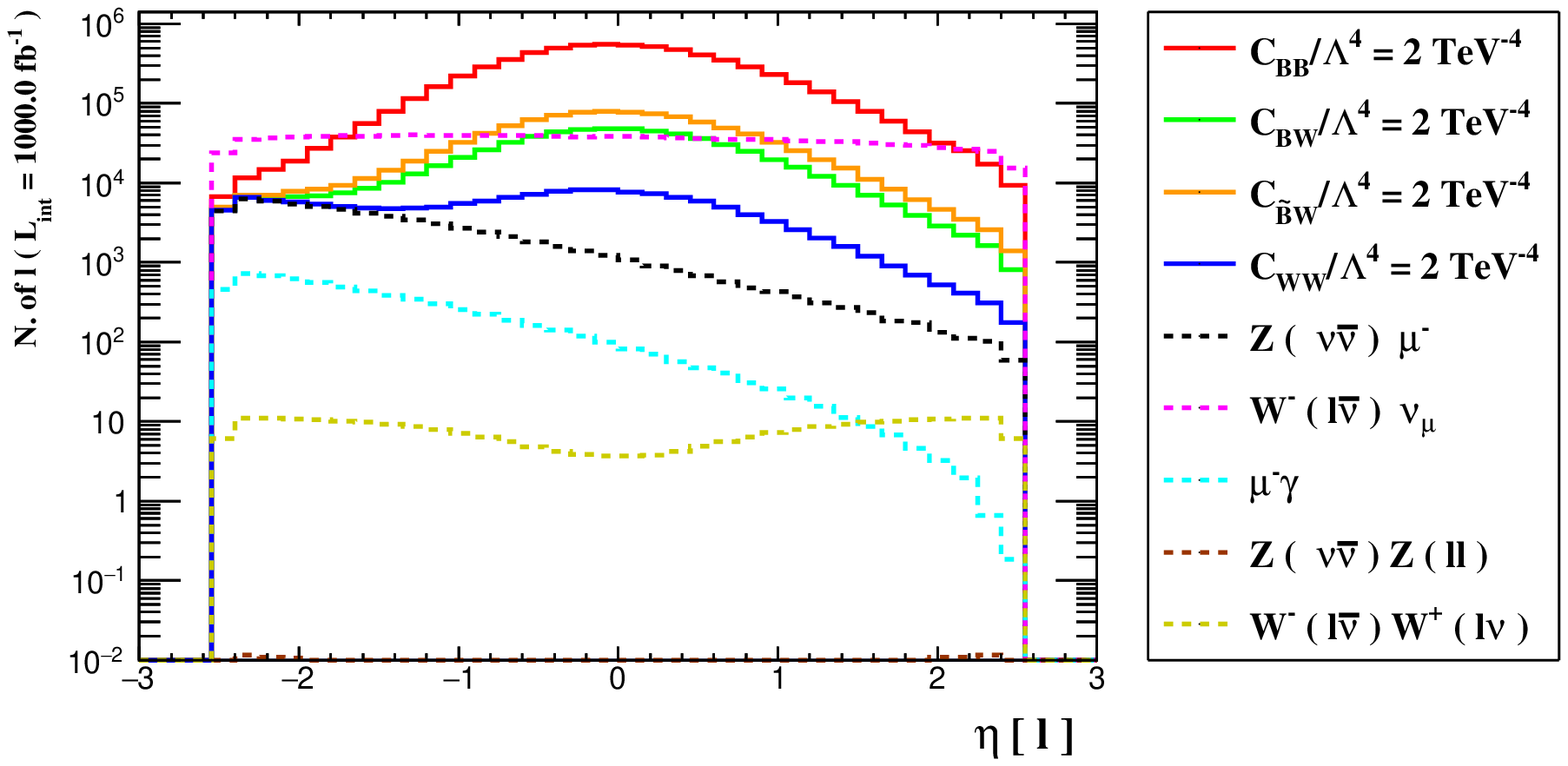}
\caption{}
\label{fig2:d}
\end{subfigure}\hfill

\caption{The distributions of charged lepton pseudo-rapidity for the signals (at 2 TeV$^{-4}$) and all relevant backgrounds processes at the (a) $\sqrt{s}=3$ TeV, (b) $\sqrt{s}=6$ TeV, (c) $\sqrt{s}=10$ TeV, (d) $\sqrt{s}=14$ TeV with ${\cal L}_{\text{int}}=1000$ fb$^{-1}$.}
\label{fig2}
\end{figure}

\begin{figure}[H]
\centering
\begin{subfigure}{0.5\linewidth}
\includegraphics[width=\linewidth]{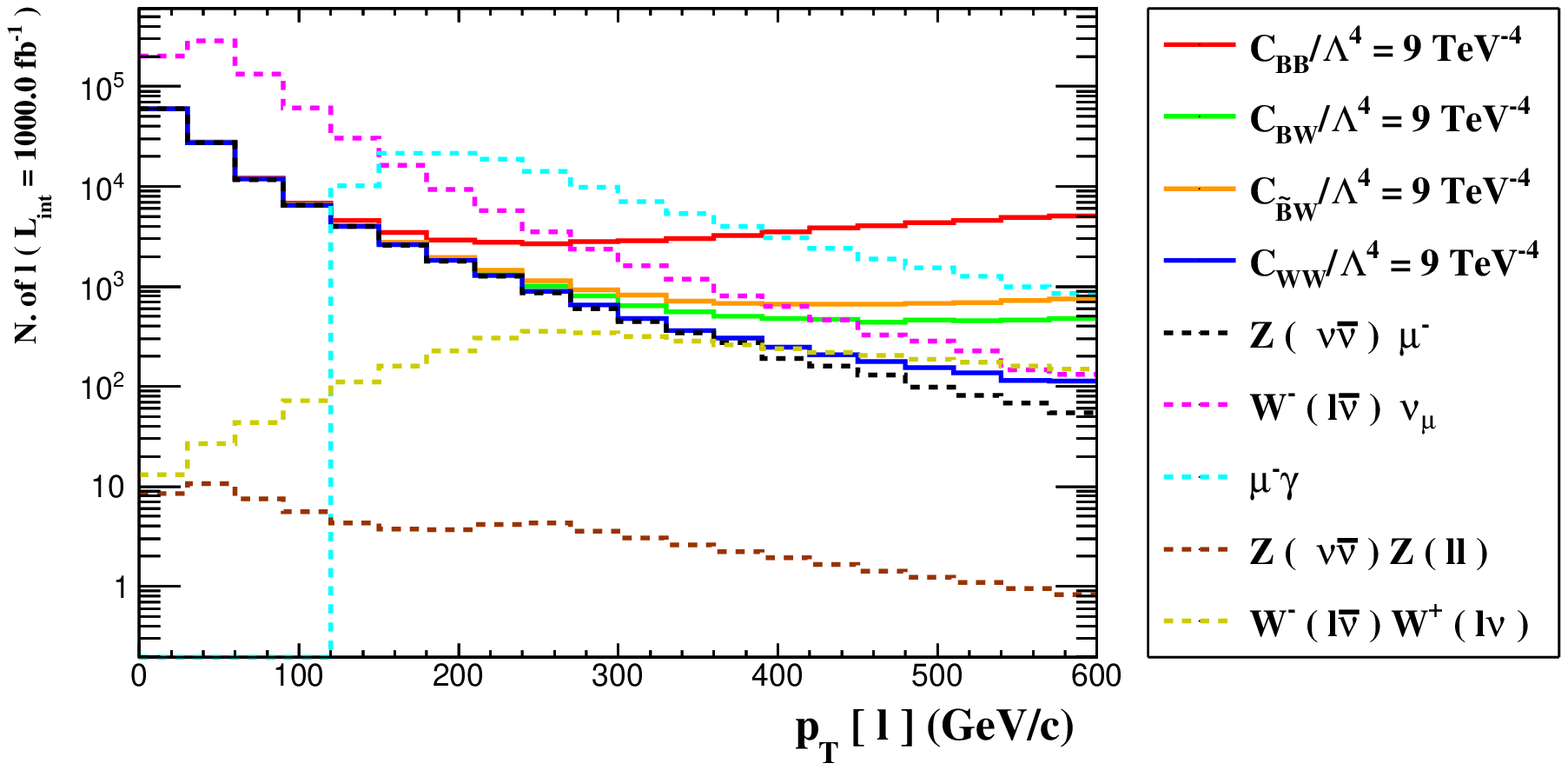}
\caption{}
\label{fig3:a}
\end{subfigure}\hfill
\begin{subfigure}{0.5\linewidth}
\includegraphics[width=\linewidth]{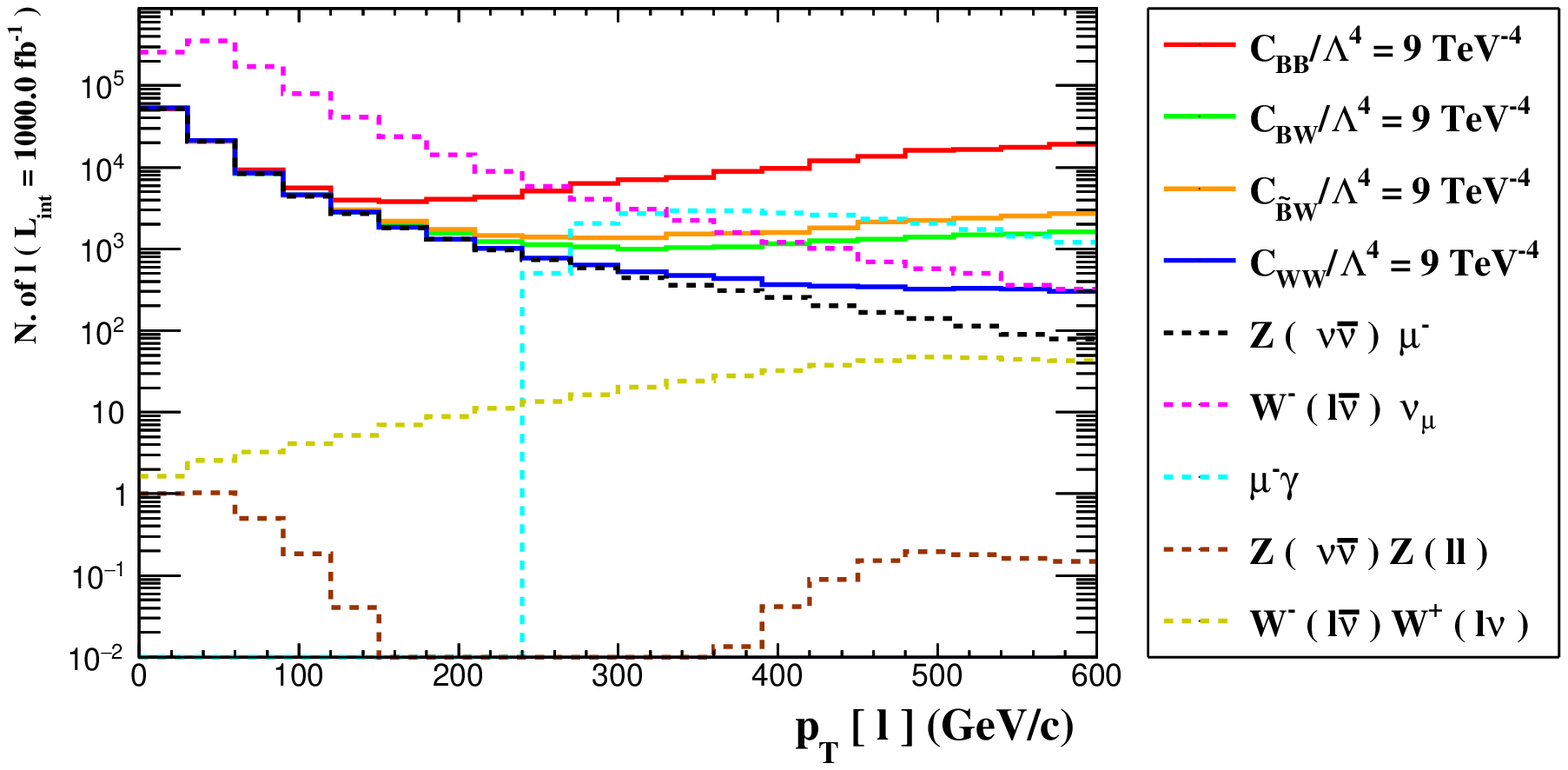}
\caption{}
\label{fig3:b}
\end{subfigure}\hfill

\begin{subfigure}{0.5\linewidth}
\includegraphics[width=\linewidth]{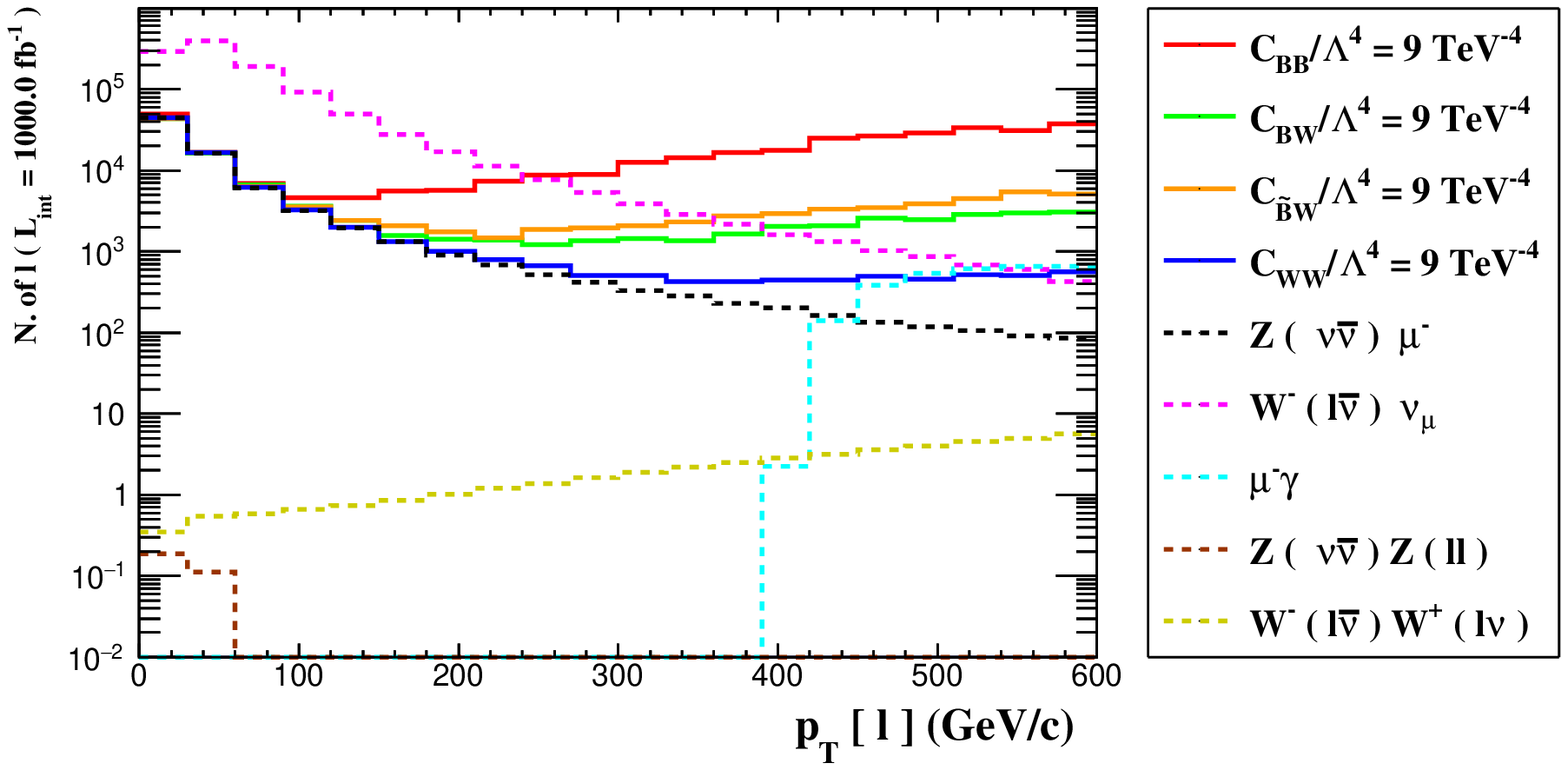}
\caption{}
\label{fig3:c}
\end{subfigure}\hfill
\begin{subfigure}{0.5\linewidth}
\includegraphics[width=\linewidth]{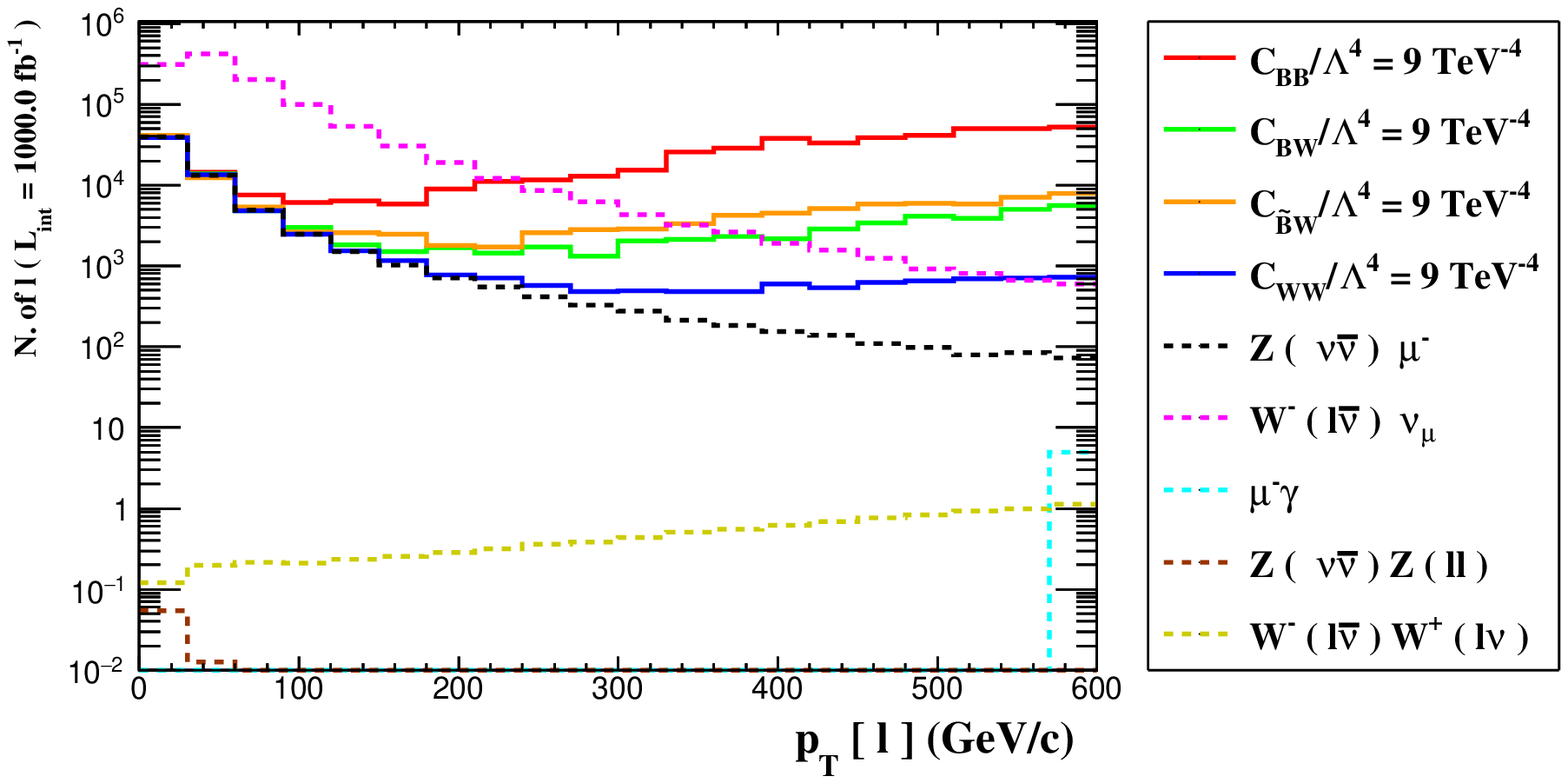}
\caption{}
\label{fig3:d}
\end{subfigure}\hfill

\caption{The distributions of charged lepton transverse momentum for the signals (at 9 TeV$^{-4}$) and all relevant backgrounds processes at the (a) $\sqrt{s}=3$ TeV, (b) $\sqrt{s}=6$ TeV, (c) $\sqrt{s}=10$ TeV, (d) $\sqrt{s}=14$ TeV with ${\cal L}_{\text{int}}=1000$ fb$^{-1}$.}
\label{fig3}
\end{figure}

\begin{figure}[H]
\centering
\begin{subfigure}{0.5\linewidth}
\includegraphics[width=\linewidth]{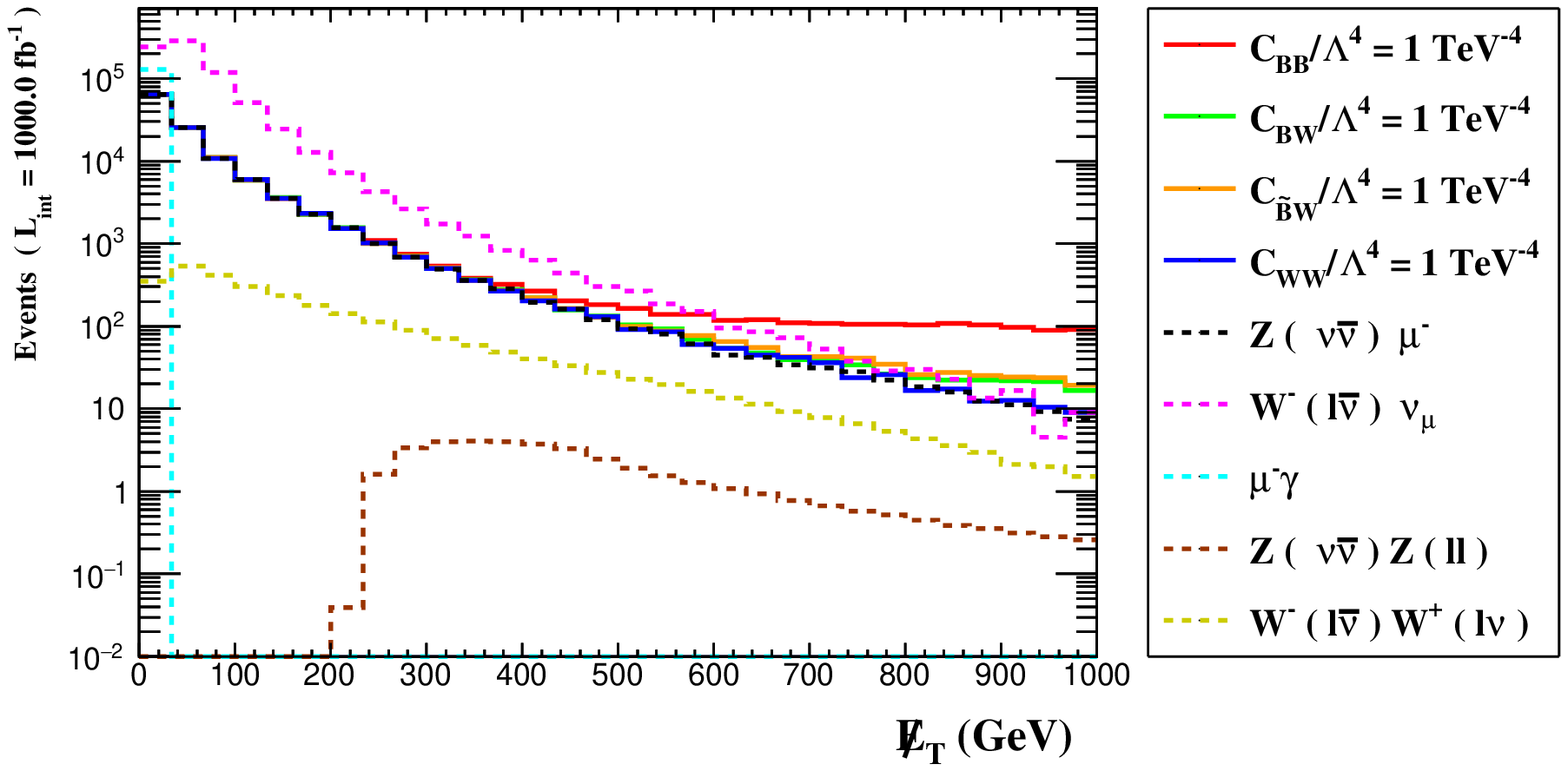}
\caption{}
\label{fig4:a}
\end{subfigure}\hfill
\begin{subfigure}{0.5\linewidth}
\includegraphics[width=\linewidth]{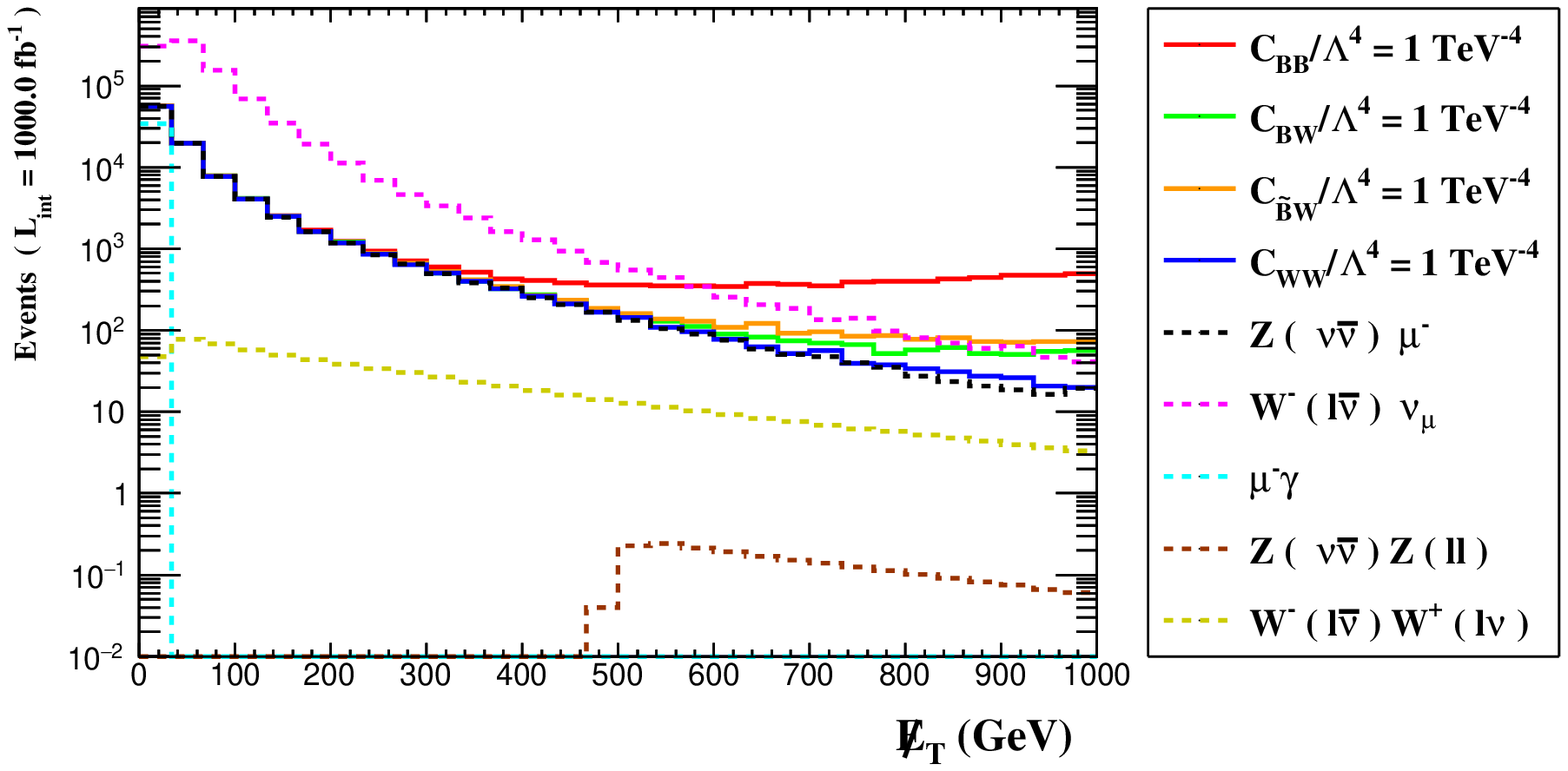}
\caption{}
\label{fig4:b}
\end{subfigure}\hfill

\begin{subfigure}{0.5\linewidth}
\includegraphics[width=\linewidth]{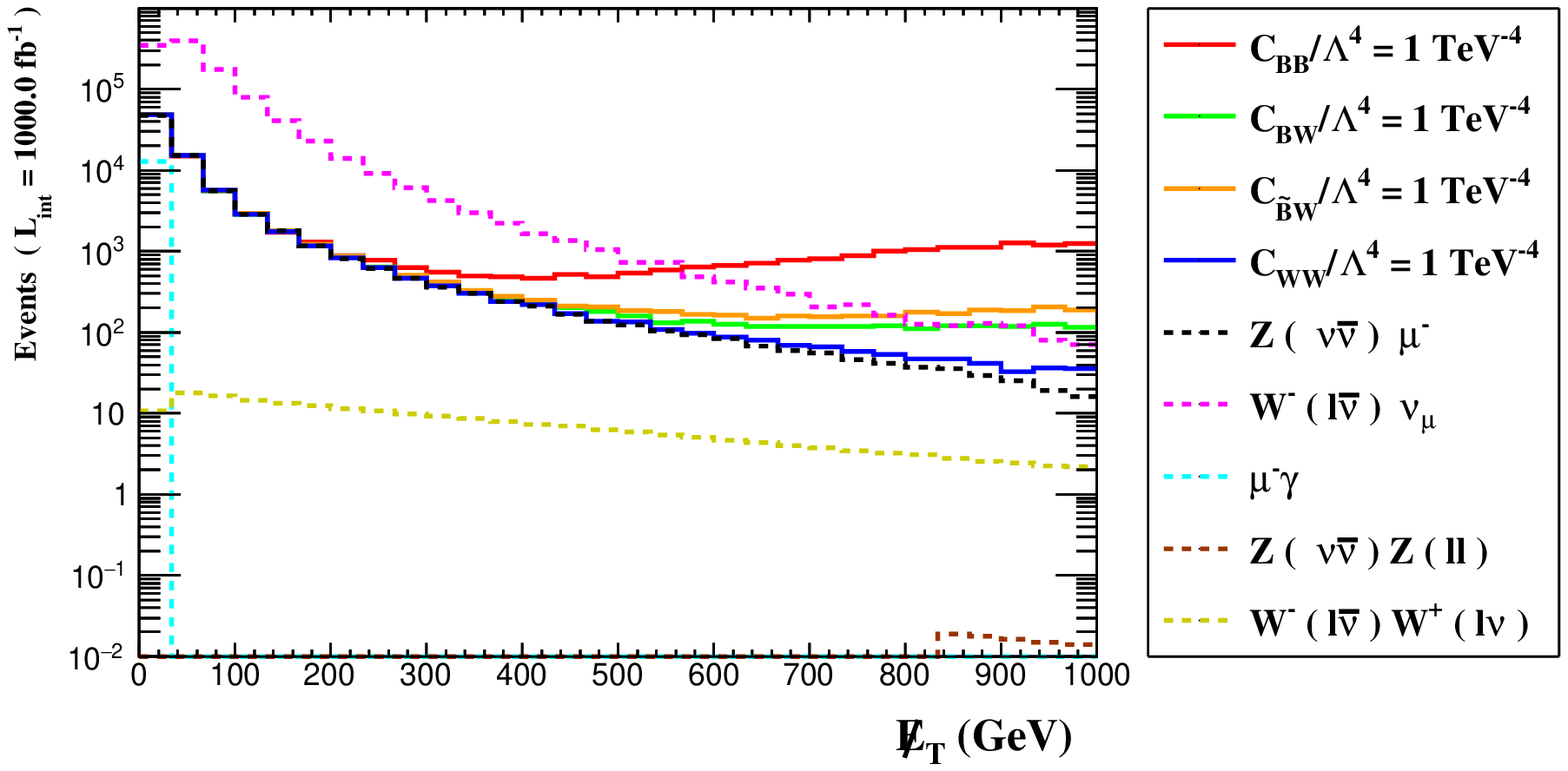}
\caption{}
\label{fig4:c}
\end{subfigure}\hfill
\begin{subfigure}{0.5\linewidth}
\includegraphics[width=\linewidth]{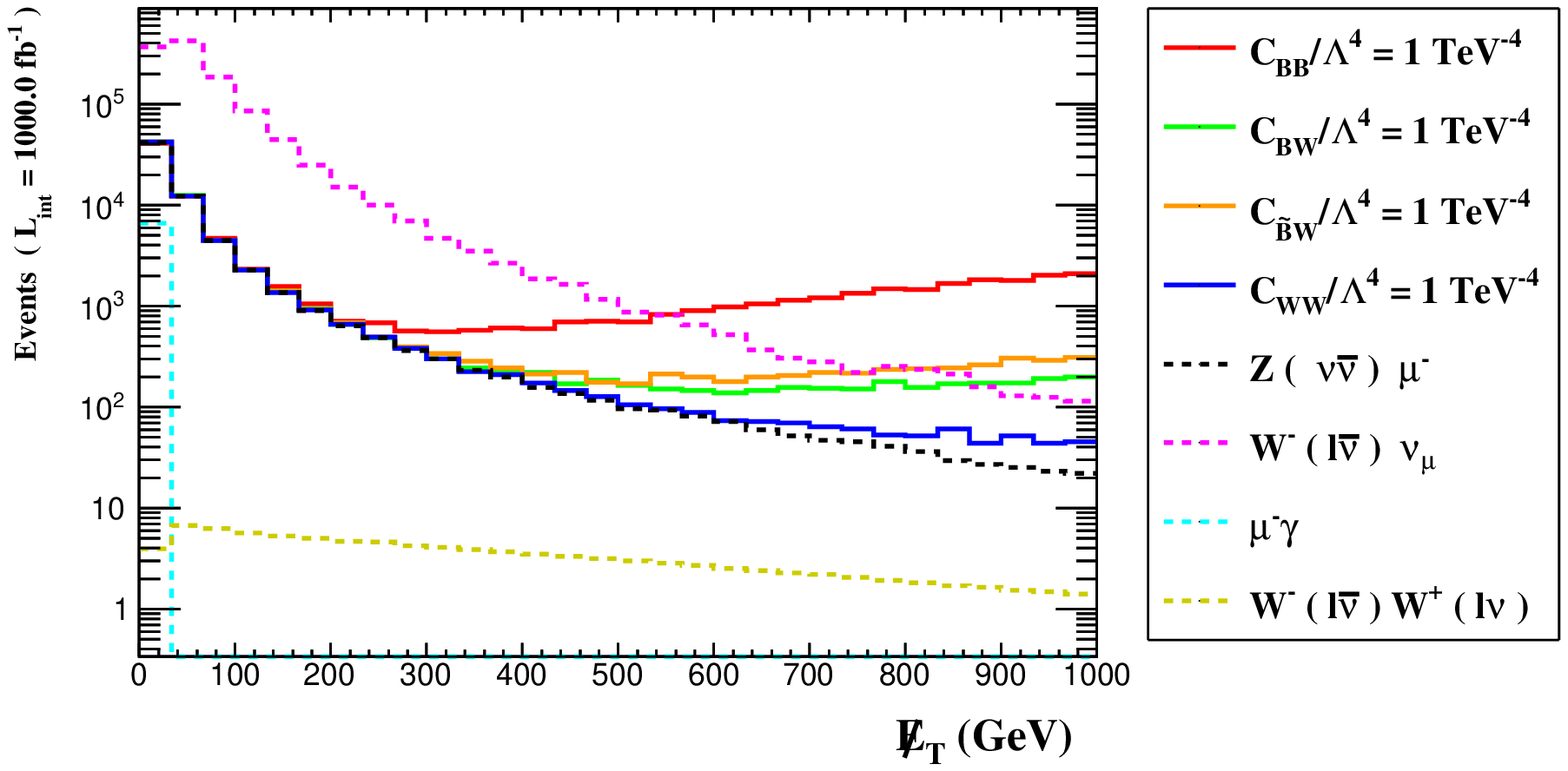}
\caption{}
\label{fig4:d}
\end{subfigure}\hfill

\caption{The distributions of transverse missing energy for the signals (at 1 TeV$^{-4}$) and all relevant backgrounds processes at the (a) $\sqrt{s}=3$ TeV, (b) $\sqrt{s}=6$ TeV, (c) $\sqrt{s}=10$ TeV, (d) $\sqrt{s}=14$ TeV with ${\cal L}_{\text{int}}=1000$ fb$^{-1}$.}
\label{fig4}
\end{figure}

In this paper, each coupling should be considered separately in order to compare the effects of anomalous couplings on the signal. The method applied for this is to assign a value to a certain anomalous coupling and set all other anomalous couplings equal to zero. The pseudo-rapidity distributions of the charged lepton $\eta^{\ell}$ for signal ($C_{BB}/{\Lambda^4}=2$ TeV$^{-4}$, $C_{BW}/{\Lambda^4}=2$ TeV$^{-4}$, $C_{\widetilde{B}W}/{\Lambda^4}=2$ TeV$^{-4}$, $C_{WW}/{\Lambda^4}=2$ TeV$^{-4}$ couplings) and for six different backgrounds are presented in Fig.~\ref{fig2} and the change in center-of-mass energies $\sqrt{s}=3$ TeV, $6$ TeV, $10$ TeV and $14$ TeV is investigated by (a)-(d), respectively. In this study, the pseudo-rapidity distributions of the charged lepton are measured using the range $|\eta^\ell|<2.5$. The pseudo-rapidity ($\eta$) is defined as $-ln[\text{tan}(\theta/2)]$ where the $\theta$ is the polar angle from the beam axis. As seen in Fig.~\ref{fig2}, the pseudo-rapidity distributions of the charged lepton for the signal peaking at $\eta=0$ (i.e. at $\theta=90^\circ$) with the increase in center-of-mass energy clearly reveal the separation from the SM background. When the pseudo-rapidity cut of $-2.5<\eta^\ell< 2.5$ is applied to charged leptons, it is expected that the efficiency of the background will decrease faster than the signal, although both the signal and background efficiencies are reduced. Thus, the signal effect becomes more pronounced, increasing the signal to background ratio. Since the $\mu^-\gamma^*$ collision with the EPA has an energy asymmetry, there is no symmetry in the forward and backward directions for the charged leptons in the process. In Fig.~\ref{fig3}, the distribution of the charged lepton transverse momentum is given for signal ($C_{BB}/{\Lambda^4}=9$ TeV$^{-4}$, $C_{BW}/{\Lambda^4}=9$ TeV$^{-4}$, $C_{\widetilde{B}W}/{\Lambda^4}=9$ TeV$^{-4}$, $C_{WW}/{\Lambda^4}=9$ TeV$^{-4}$ couplings) and for six different backgrounds. In these distributions, which are ordered from (a) to (d) according to the center-of-mass energies, it is seen that there are varying deviations between the signals and the SM background due to the increase in the center-of-mass energy. The distribution of transverse missing energy given from (a) to (d) with the center-of-mass energies is examined in Fig.~\ref{fig4} for signal ($C_{BB}/{\Lambda^4}=1$ TeV$^{-4}$, $C_{BW}/{\Lambda^4}=1$ TeV$^{-4}$, $C_{\widetilde{B}W}/{\Lambda^4}=1$ TeV$^{-4}$, $C_{WW}/{\Lambda^4}=1$ TeV$^{-4}$ couplings) and for six different backgrounds. There are different deviations between the signals and the SM background according to the center-of-mass energy. However, the deviations between the signals and the backgrounds seen in Figs.~\ref{fig2}-\ref{fig4} indicate the necessity of improving the signals by suppressing the background with $\eta^{\ell}$, $p^\ell_T$ and $\slashed{E}_T$ cuts.

The cross-sections are given in Table~\ref{tab4} to examine the effects on the signals and relevant backgrounds after the basic cuts and selected cuts. Here, $S_i$ is the cross section of the signal and $B_i$ is the cross section of the backgrounds. It indicates that the basic cut is applied if the index $i=1$ and that the selected cut is applied if the index $i=2$. It is understood that the selected cuts suppress the relevant background according to the low values of the $B_2/B_1$ percentage and do not cause signal loss according to the high values of the $S_2/S_1$ percentage. If the total backgrounds $B_{tot_1}$ and $B_{tot_2}$ after the basic cuts and selected cuts are analyzed as the signal to total background ratio, the ratio after the selected cuts improves significantly compared to the ratio after the basic cuts. It is seen that the selected cuts suppress the relevant backgrounds and thus the signals are much more prominent than these backgrounds.

\begin{table}[H]
\centering
\caption{Cross-sections for the signals and the backgrounds according to the basic and selected cuts at 14 TeV muon collider.}
\label{tab4}
\begin{ruledtabular}
\begin{tabular}{llclcc}
\multirow{3}{*}{Signals} & Cross-sections & \multirow{3}{*}{$S_1/B_{tot_1}$} & Cross-sections & \multirow{3}{*}{$S_2/B_{tot_2}$} & $S_2/S_1$\\
 & with basic cuts & & with selected cuts & & [$\%$]\\ 
 & $S_1$ (pb) & & $S_2$ (pb) & & \\ \hline
$C_{BB}/{\Lambda^4}=3$ TeV$^{-4}$ & 15.542 & 12.250 & 15.486 & 242.499 & $\%$99.6\\ 
$C_{BW}/{\Lambda^4}=3$ TeV$^{-4}$ & 1.383 & 1.090 & 1.318 & 20.639 & $\%$95.3\\ 
$C_{\widetilde{B}W}/{\Lambda^4}=3$ TeV$^{-4}$ & 2.239 & 1.765 & 2.169 & 33.965 & $\%$96.9\\ 
$C_{WW}/{\Lambda^4}=3$ TeV$^{-4}$ & 0.259 & 0.204 & 0.197 & 3.085 & $\%$76.1\\ 
\hline
\multirow{3}{*}{Backgrounds} & Cross-sections & & Cross-sections & & $B_2/B_1$\\
 & with basic cuts & & with selected cuts & & [$\%$]\\ 
 & $B_1$ (pb) & & $B_2$ (pb) & & \\ \hline
$Z(\nu\bar{\nu})\mu^-$ & 0.06610 & & 0.00363 & & $\%$5.5\\ 
$W^-(\ell^- \bar{\nu})\nu_\mu$ & 1.18588 & & 0.05341 & & $\%$4.5\\ 
$\mu^-\gamma$ & 0.01662 & & 0.00673 & & $\%$40.5\\ 
$Z(\nu\bar{\nu})Z(\ell^- \ell^+)$ & 6.75$\times10^{-8}$ & & 1.27$\times10^{-8}$ & & $\%$18.8\\ 
$W^-(\ell^- \bar{\nu})W^+(\ell^+ \nu)$ & 0.00013 & & 0.00009 & & $\%$69.2\\ 
  & $B_{tot_1}=1.26873$ & & $B_{tot_2}=0.06386$ & & \\
\end{tabular}
\end{ruledtabular}
\end{table}

\begin{figure}[H]
\centering
\begin{subfigure}{0.5\linewidth}
\includegraphics[width=\linewidth]{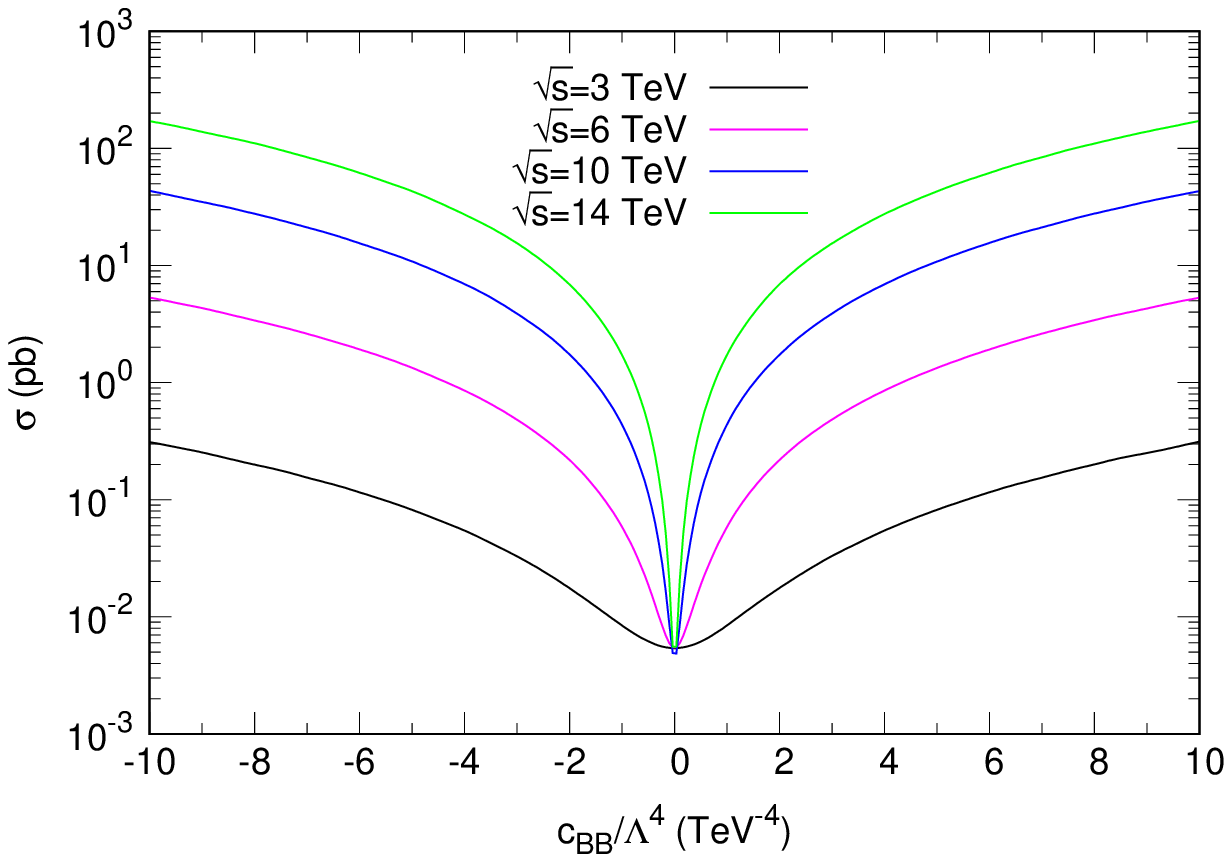}
\caption{}
\label{fig5:a}
\end{subfigure}\hfill
\begin{subfigure}{0.5\linewidth}
\includegraphics[width=\linewidth]{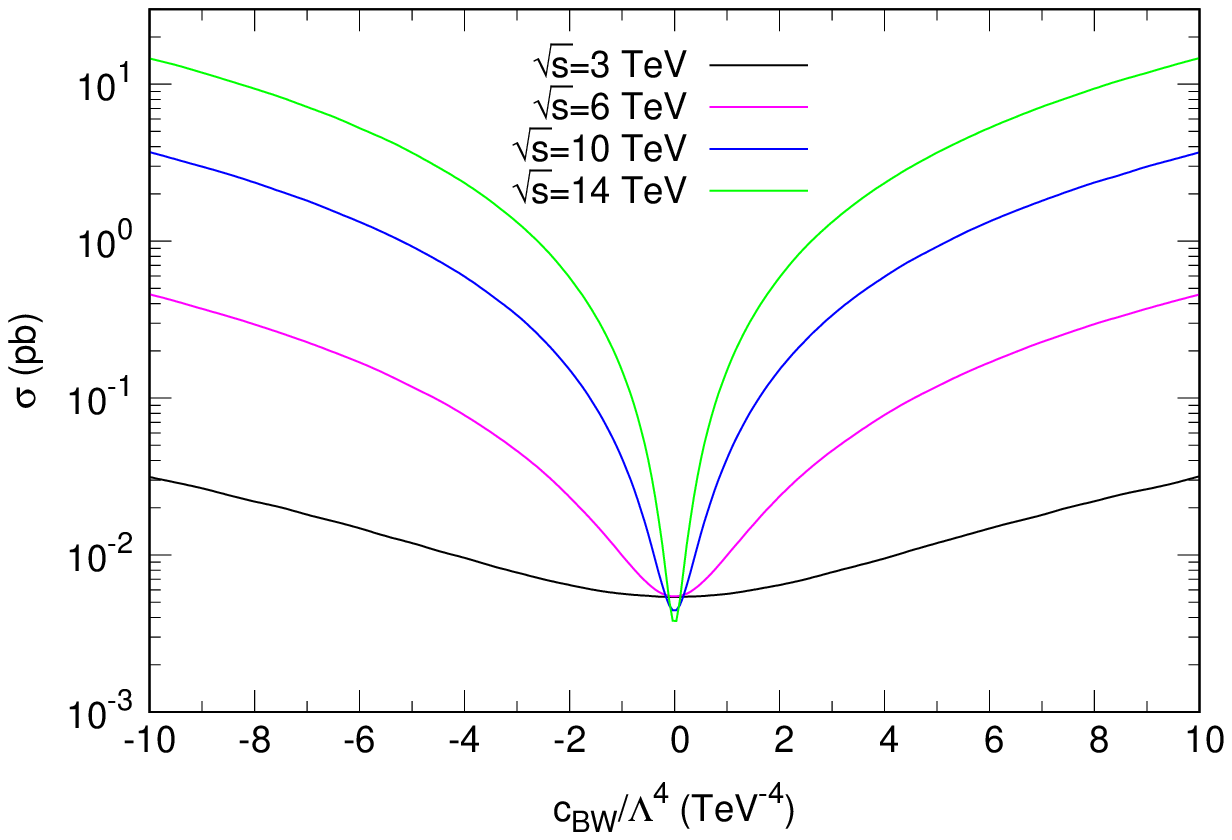}
\caption{}
\label{fig5:b}
\end{subfigure}\hfill

\begin{subfigure}{0.5\linewidth}
\includegraphics[width=\linewidth]{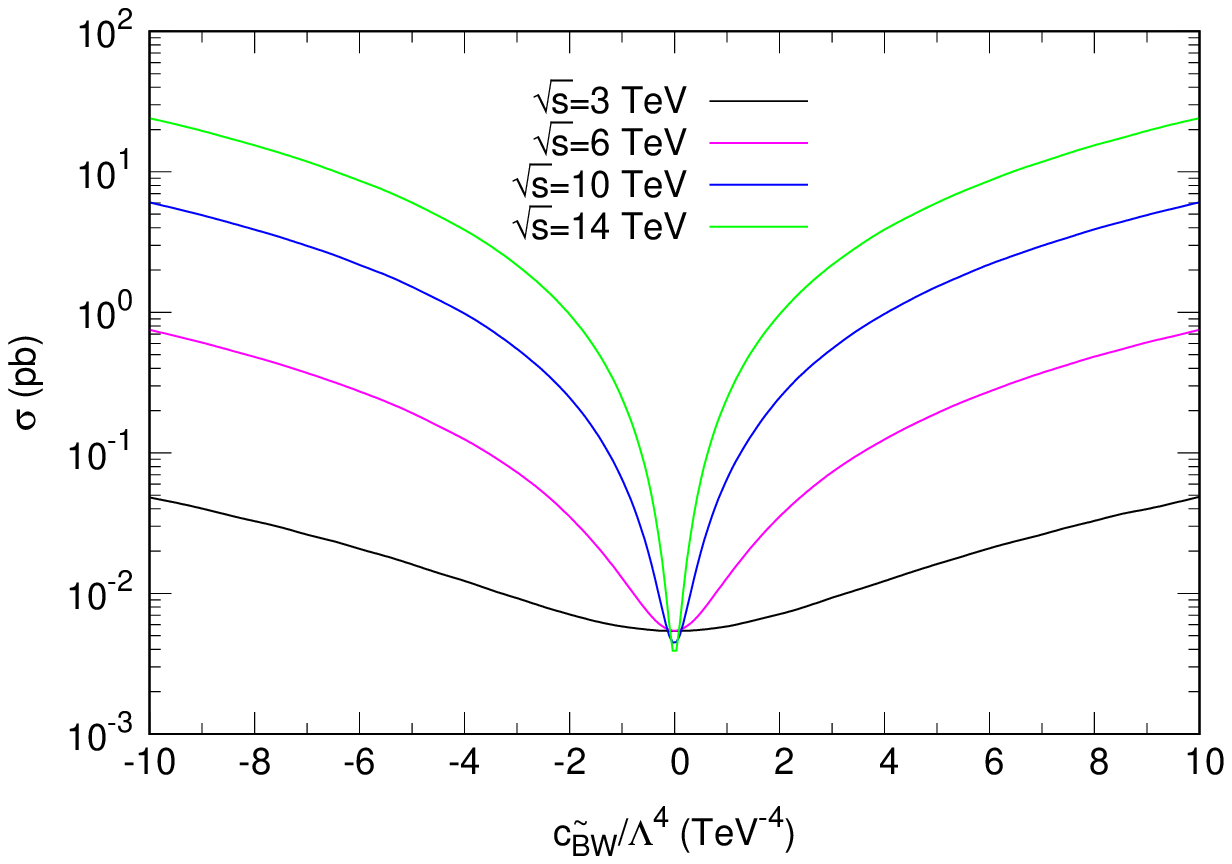}
\caption{}
\label{fig5:c}
\end{subfigure}\hfill
\begin{subfigure}{0.5\linewidth}
\includegraphics[width=\linewidth]{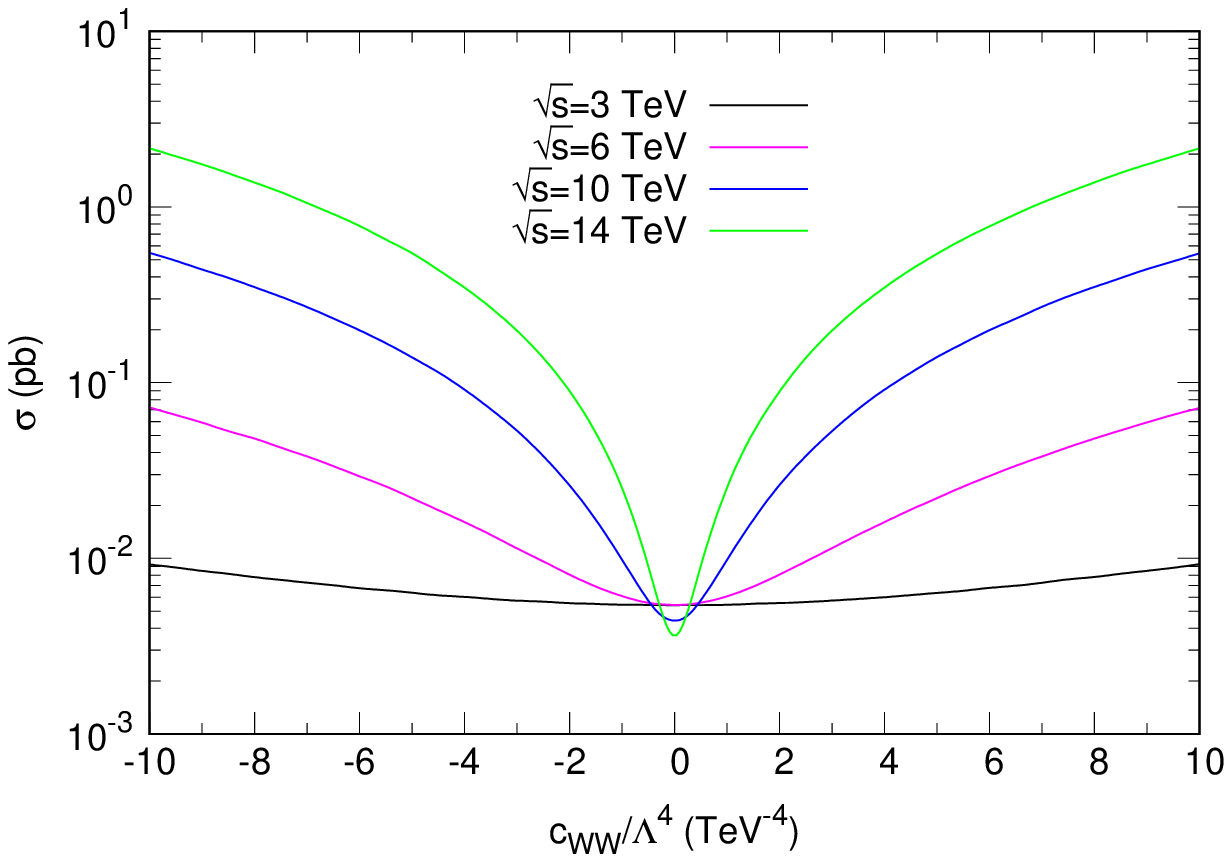}
\caption{}
\label{fig5:d}
\end{subfigure}\hfill

\caption{The total cross sections of the process $\mu^+\mu^-\,\rightarrow\,\mu^+\gamma^*\mu^-\,\rightarrow\,\mu^+ Z(\nu\bar{\nu})\mu^-$ as a function of the anomalous (a) $C_{BB}/{\Lambda^4}$, (b) $C_{BW}/{\Lambda^4}$, (c) $C_{\widetilde{B}W}/{\Lambda^4}$, (d) $C_{WW}/{\Lambda^4}$ couplings for center-of-mass energies of $\sqrt{s}=3$ TeV, 6 TeV, 10 TeV, 14 TeV.}
\label{fig5}
\end{figure}

The total cross-sections of the process $\mu^+\mu^-\,\rightarrow\,\mu^+\gamma^*\mu^-\,\rightarrow\,\mu^+ Z(\nu\bar{\nu})\mu^-$ as a function of anomalous $C_{BB}/{\Lambda^4}$, $C_{BW}/{\Lambda^4}$, $C_{\widetilde{B}W}/{\Lambda^4}$ and $C_{WW}/{\Lambda^4}$ couplings in the muon collider with $\sqrt{s}=3$ TeV, $6$ TeV, $10$ TeV, $14$ TeV are presented in Fig.~\ref{fig5}. The total cross-section corresponding to the function of an anomalous coupling has studied by fixing the other three couplings to zero. In addition, the selected cuts have applied in the analysis of these total cross-sections. It is seen that the total cross-section of each anomalous coupling increases as the center-of-mass energy of the muon collider increases.

\begin{figure}[H]
\centering
\includegraphics[scale=0.7]{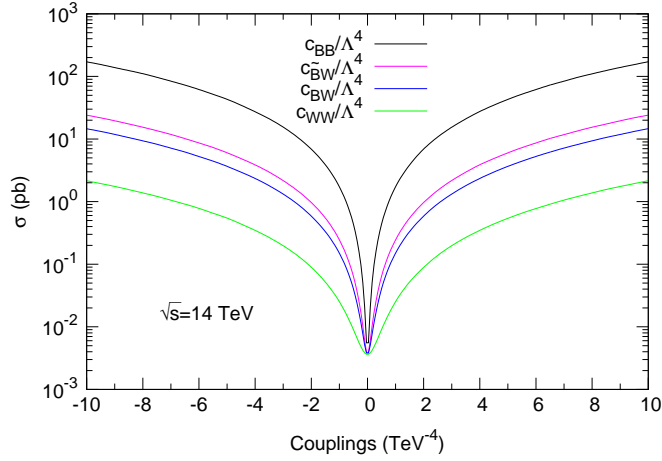}
\caption{The total cross sections of the process $\mu^+\mu^-\,\rightarrow\,\mu^+\gamma^*\mu^-\,\rightarrow\,\mu^+ Z(\nu\bar{\nu})\mu^-$ as a function of the anomalous couplings at $\sqrt{s}=14$ TeV.
\label{fig6}}
\end{figure}

In Fig.~\ref{fig6}, the total cross-sections of anomalous $C_{BB}/{\Lambda^4}$, $C_{BW}/{\Lambda^4}$, $C_{\widetilde{B}W}/{\Lambda^4}$ and $C_{WW}/{\Lambda^4}$ couplings in the muon collider with $\sqrt{s}=14$ TeV are compared with each other. It is seen that the anomalous $C_{BB}/{\Lambda^4}$ coupling has the highest cross-section and they are ordered from highest to lowest as $C_{\widetilde{B}W}/{\Lambda^4}$, $C_{BW}/{\Lambda^4}$ and $C_{WW}/{\Lambda^4}$.

\section{Sensitivities on the anomalous neutral triple gauge couplings} \label{Sec3}

The 95$\%$ C.L. limit is calculated using a $\chi^2$ test with systematic errors to probe the sensitivities of anomalous $C_{BB}/{\Lambda^4}$, $C_{BW}/{\Lambda^4}$, $C_{\widetilde{B}W}/{\Lambda^4}$ and $C_{WW}/{\Lambda^4}$ couplings. $\chi^2$ test is defined by

\begin{eqnarray}
\label{eq.21} 
\chi^2=\left(\frac{\sigma_{B_{tot}}-\sigma_{NP}}{\sigma_{B_{tot}}\sqrt{\left(\delta_{st}\right)^2+\left(\delta_{sys}\right)^2}}\right)^2
\end{eqnarray}

{\raggedright where $\sigma_{B_{tot}}$ is only the cross section of total background and $\sigma_{NP}$ is the cross section in the presence of both new physics beyond the SM and total background. $\delta_{st}=\frac{1}{\sqrt {N_{B_{tot}}}}$ and $\delta_{sys}$ are the statistical error and the systematic error, respectively. The number of events in the total backgrounds is described as $N_{B_{tot}}={\cal L}_{\text{int}} \times \sigma_{B_{tot}}$, where ${\cal L}_{\text{int}}$ is the integrated luminosity. The systematic uncertainties are included in the statistical analysis of the $\chi^2$ test for many reasons \cite{Khoriauli:2008xza}. In this study, the systematic uncertainties of $0\%$, $3\%$ and $5\%$ in the anomalous couplings are discussed.}

The 95$\%$ C.L. limits of anomalous $C_{BB}/{\Lambda^4}$, $C_{BW}/{\Lambda^4}$, $C_{\widetilde{B}W}/{\Lambda^4}$ and $C_{WW}/{\Lambda^4}$ couplings through process $\mu^+\mu^-\,\rightarrow\,\mu^+\gamma^*\mu^-\,\rightarrow\,\mu^+ Z(\nu\bar{\nu})\mu^-$ at the muon collider are investigated in Tables~\ref{tab5}-\ref{tab8} without and with systematic error $3\%$, $5\%$. The $95\%$ C.L. limits in Tables~\ref{tab5}-\ref{tab8} are obtained by solving $\chi^2=3.84$ for Eq.~(\ref{eq.21}). Sensitivities are evaluated for both basic and selected cuts using center-of-mass energies and integrated luminosities of the muon collider given in Table~\ref{tab1}.

\begin{table}[H]
\caption{The 95\% C.L. limits on the anomalous $C_{BB}/{\Lambda^4}$ coupling with respect to center-of-mass energies and cuts for systematic errors of $0\%$, $3\%$ and $5\%$.}
\label{tab5}
\begin{ruledtabular}
\begin{tabular}{ccccc}
\multicolumn{2}{c}{} & \multicolumn{3}{c}{$C_{BB}/{\Lambda^4}$ (TeV$^{-4}$)} \\
\hline
$\sqrt{s}$ & Cuts & $\delta_{sys}=0\%$ & $\delta_{sys}=3\%$ & $\delta_{sys}=5\%$ \\ 
\hline \hline
\multirow{2}{*}{3 TeV} 
 & Basic & [-0.80008; 0.78930] & [-4.35957; 4.35199] & [-5.62634; 5.62098]\\ 
 & Selected & [-0.50315; 0.50441] & [-1.73497; 1.73613] & [-2.23742; 2.23850]\\ \hline
\multirow{2}{*}{6 TeV} 
 & Basic & [-0.13972; 0.13760] & [-1.10028; 1.09826] & [-1.42008; 1.41812]\\ 
 & Selected & [-0.07659; 0.06584] & [-0.29384; 0.28310] & [-0.37753; 0.36681]\\ \hline
\multirow{2}{*}{10 TeV}  
 & Basic & [-0.03857; 0.04068] & [-0.40254; 0.40465] & [-0.51997; 0.52209]\\ 
 & Selected & [-0.02349; 0.01546] & [-0.09748; 0.08945] & [-0.12460; 0.11657]\\ \hline
\multirow{2}{*}{14 TeV} 
 & Basic & [-0.01919; 0.01490] & [-0.20968; 0.20539] & [-0.27006; 0.26578]\\  
 & Selected & [-0.01026; 0.00636] & [-0.04902; 0.04512] & [-0.06269; 0.05879]\\ 
\end{tabular}
\end{ruledtabular}
\end{table}

\begin{table}[H]
\caption{Same as in Table~\ref{tab5}, but for the anomalous $C_{BW}/{\Lambda^4}$ coupling.}
\label{tab6}
\begin{ruledtabular}
\begin{tabular}{ccccc}
\multicolumn{2}{c}{} & \multicolumn{3}{c}{$C_{BW}/{\Lambda^4}$ (TeV$^{-4}$)} \\
\hline
$\sqrt{s}$ & Cuts & $\delta_{sys}=0\%$ & $\delta_{sys}=3\%$ & $\delta_{sys}=5\%$ \\ 
\hline \hline
\multirow{2}{*}{3 TeV} 
 & Basic & [-2.98930; 3.12269] & [-13.9406; 13.8188] & [-16.8060; 16.6318]\\ 
 & Selected & [-1.72510; 1.73537] & [-5.94831; 5.95581] & [-7.66432; 7.66984]\\ \hline
\multirow{2}{*}{6 TeV} 
 & Basic & [-0.46682; 0.48030] & [-3.75051; 3.76149] & [-4.84594; 4.85522]\\ 
 & Selected & [-0.25097; 0.23650] & [-0.99668; 0.98239] & [-1.28381; 1.26965]\\ \hline
\multirow{2}{*}{10 TeV}  
 & Basic & [-0.14233; 0.12953] & [-1.38957; 1.37735] & [-1.79190; 1.78006]\\ 
 & Selected & [-0.07196; 0.05922] & [-0.32636; 0.31365] & [-0.41933; 0.40664]\\ \hline
\multirow{2}{*}{14 TeV} 
 & Basic & [-0.06201; 0.05422] & [-0.71536; 0.70769] & [-0.92236; 0.91476]\\ 
 & Selected & [-0.02482; 0.03053] & [-0.15749; 0.16319] & [-0.20410; 0.20980]\\ 
\end{tabular}
\end{ruledtabular}
\end{table}

\begin{table}[H]
\caption{Same as in Table~\ref{tab5}, but for the anomalous $C_{\widetilde{B}W}/{\Lambda^4}$ coupling.}
\label{tab7}
\begin{ruledtabular}
\begin{tabular}{ccccc}
\multicolumn{2}{c}{} & \multicolumn{3}{c}{$C_{\widetilde{B}W}/{\Lambda^4}$ (TeV$^{-4}$)} \\
\hline
$\sqrt{s}$ & Cuts & $\delta_{sys}=0\%$ & $\delta_{sys}=3\%$ & $\delta_{sys}=5\%$ \\ 
\hline \hline
\multirow{2}{*}{3 TeV} 
 & Basic & [-2.27698; 2.23342] & [-11.4351; 11.4573] & [-14.1927; 14.2394]\\ 
 & Selected & [-1.35194; 1.34093] & [-4.64156; 4.62956] & [-5.98110; 5.96838]\\ \hline
\multirow{2}{*}{6 TeV} 
 & Basic & [-0.36744; 0.36919] & [-2.92076; 2.92239] & [-3.77222; 3.77377]\\ 
 & Selected & [-0.19255; 0.18721] & [-0.77381; 0.76847] & [-0.99761; 0.99229]\\ \hline
\multirow{2}{*}{10 TeV}  
 & Basic & [-0.11403; 0.09788] & [-1.08438; 1.06868] & [-1.39748; 1.38207]\\ 
 & Selected & [-0.05130; 0.05033] & [-0.24953; 0.24855] & [-0.32191; 0.32093]\\ \hline
\multirow{2}{*}{14 TeV} 
 & Basic & [-0.03968; 0.05152] & [-0.54894; 0.56074] & [-0.71038; 0.72215]\\ 
 & Selected & [-0.01830; 0.02510] & [-0.12148; 0.12829] & [-0.15778; 0.16458]\\ 
\end{tabular}
\end{ruledtabular}
\end{table}

\begin{table}[H]
\caption{Same as in Table~\ref{tab5}, but for the anomalous $C_{WW}/{\Lambda^4}$ coupling.}
\label{tab8}
\begin{ruledtabular}
\begin{tabular}{ccccc}
\multicolumn{2}{c}{} & \multicolumn{3}{c}{$C_{WW}/{\Lambda^4}$ (TeV$^{-4}$)} \\
\hline
$\sqrt{s}$ & Cuts & $\delta_{sys}=0\%$ & $\delta_{sys}=3\%$ & $\delta_{sys}=5\%$ \\ 
\hline \hline
\multirow{2}{*}{3 TeV} 
 & Basic & [-8.18675; 8.00071] & [-35.8366; 30.9190] & [-42.2103; 36.6817]\\ 
 & Selected & [-4.52178; 4.53890] & [-15.3627; 15.2800] & [-19.5881; 19.4449]\\ \hline
\multirow{2}{*}{6 TeV} 
 & Basic & [-1.30918; 1.27245] & [-9.77263; 9.76235] & [-12.2887; 12.2904]\\ 
 & Selected & [-0.64557; 0.62702] & [-2.59265; 2.57541] & [-3.34211; 3.32579]\\ \hline
\multirow{2}{*}{10 TeV}  
 & Basic & [-0.36063; 0.34319] & [-3.59278; 3.58033] & [-4.63648; 4.62740]\\ 
 & Selected & [-0.17759; 0.16369] & [-0.84245; 0.82876] & [-1.08521; 1.07166]\\ \hline
\multirow{2}{*}{14 TeV} 
 & Basic & [-0.15065; 0.15109] & [-1.85110; 1.85159] & [-2.38992; 2.39044]\\ 
 & Selected & [-0.06981; 0.07387] & [-0.41624; 0.42028] & [-0.53788; 0.54191]\\ 
\end{tabular}
\end{ruledtabular}
\end{table}

\begin{figure}[H]
\centering
\includegraphics[scale=0.7]{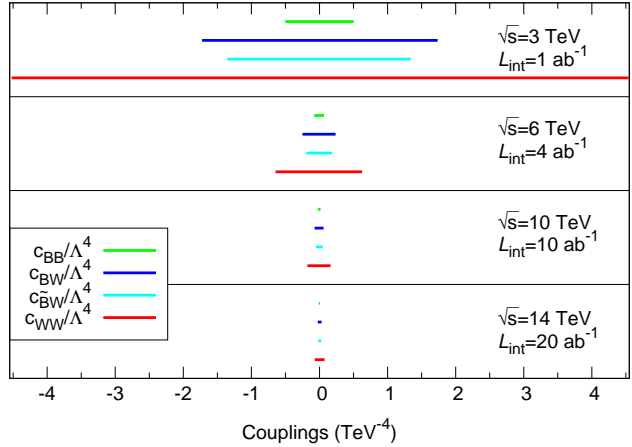}
\caption{Comparison of the sensitivities for the anomalous couplings through the process $\mu^+\mu^-\,\rightarrow\,\mu^+\gamma^*\mu^-\,\rightarrow\,\mu^+ Z(\nu\bar{\nu})\mu^-$ at the muon collider with center-of-mass energies $\sqrt{s}=3$, $6$, $10$, $14$ TeV and integrated luminosities ${\cal L}_{\text{int}}=1$, $4$, $10$, $20$ ab$^{-1}$.
\label{fig7}}
\end{figure}

In order to easily compare the sensitivities of anomalous couplings between scenarios of muon collider, 95$\%$ C.L. limits without systematic error in selected cuts given in Tables~\ref{tab5}-\ref{tab8} are examined in Fig.~\ref{fig7}. The comparison reveals that the increase in center-of-mass energies and integrated luminosities in the muon collider, as well as the use of selected cuts, increase the sensitivity of the limits.

The most sensitive limits appear to be obtained from the selected cuts at the 14 TeV muon collider. These most sensitive limits without systematic error are as follows:

\begin{eqnarray}
\label{eq.22} 
C_{BB}/{\Lambda^4}=[-0.01026; 0.00636]\,\text{TeV}^{-4}\,,
\end{eqnarray}
\begin{eqnarray}
\label{eq.23} 
C_{BW}/{\Lambda^4}=[-0.02482; 0.03053]\,\text{TeV}^{-4}\,,
\end{eqnarray}
\begin{eqnarray}
\label{eq.24} 
C_{\widetilde{B}W}/{\Lambda^4}=[-0.01830; 0.02510]\,\text{TeV}^{-4}\,,
\end{eqnarray}
\begin{eqnarray}
\label{eq.25} 
C_{WW}/{\Lambda^4}=[-0.06981; 0.07387]\,\text{TeV}^{-4}\,.
\end{eqnarray}

\section{Conclusions} \label{Sec4}

The photon-induced process $\mu^+\mu^-\,\rightarrow\,\mu^+\gamma^*\mu^-\,\rightarrow\,\mu^+ Z(\nu\bar{\nu})\mu^-$ has been preferred to probe $Z\gamma\gamma$ and $ZZ\gamma$ aNTGC in the multi-TeV muon collider. In the analysis, cuts such as the charged lepton pseudo-rapidity, the charged lepton transverse momentum and the transverse missing energy are applied to separate the signal and the relevant SM background. The distinction between signal and background is tested by discussing the distributions of these cuts according to the number of events. In addition, it is concluded that the signal to background ratio increase with the selected cuts. Total cross-sections of the process for the anomalous couplings are presented at different scenarios of muon collider.

The significance of the study is demonstrated by obtaining the sensitivities of $95\%$ C.L. limits for dimension-eight CP-conserving $C_{\widetilde{B}W}/{\Lambda^4}$ coupling and CP-violating $C_{BB}/{\Lambda^4}$, $C_{BW}/{\Lambda^4}$, $C_{WW}/{\Lambda^4}$ couplings. Sensitivity limits are calculated for both basic and selected cuts for $0\%$, $3\%$ and $5\%$ systematic uncertainties in muon collider scenarios with different center-of-mass energies and integrated luminosities. In this way, it has been checked that the sensitivity of the limits increases with the selected cuts, and more importantly, it is proven that the sensitivity of the limits increased significantly as center-of-mass energies and integrated luminosities increase of the muon collider.

If the most sensitive limits in Eqs.~\ref{eq.22}-\ref{eq.25} are compared with the latest and sensitive experimental results from the LHC, there is an improvement of approximately 31, 24, 52, and 32 times in sensitivities of the aNTGC couplings, respectively. Except for the muon collider scenario with center-of-mass energy of 3 TeV and integrated luminosities of 1 ab$^{-1}$, the other three scenarios seem to have sensitive limits above the experimental limits at the LHC. 

The High Luminosity Large Hadron Collider (HL-LHC) with the same center-of-mass energy as the 14 TeV muon collider, which gives the most sensitive limits in our study, will be operational after 2029 according to the latest plans. Phenomenological studies of aNTGC for the $Z\gamma\gamma$ and $ZZ\gamma$ vertices at the HL-LHC have been performed through the processes $pp \rightarrow \ell^+ \ell^- \gamma$ \cite{Senol:2019ybv} and $pp \rightarrow \nu\nu\gamma$ \cite{Senol:2020hbh}. Sensitivity limits at $95\%$ C.L. in Ref.~\cite{Senol:2019ybv} have been found to be $[-1.47; 1.47]\,\text{TeV}^{-4}$ and $[-0.86; 0.86]\,\text{TeV}^{-4}$ for anomalous $C_{BB}/{\Lambda^4}$ coupling and $[-1.88; 1.88]\,\text{TeV}^{-4}$ and $[-1.14; 1.14]\,\text{TeV}^{-4}$ for anomalous $C_{\widetilde{B}W}/{\Lambda^4}$ coupling with an integrated luminosity of 0.3 ab$^{-1}$ and 3 ab$^{-1}$, respectively, at HL-LHC. The $95\%$ C.L. limits of anomalous $C_{BB}/{\Lambda^4}$, $C_{BW}/{\Lambda^4}$, $C_{\widetilde{B}W}/{\Lambda^4}$ and $C_{WW}/{\Lambda^4}$ couplings have been obtained as $[-0.21; 0.21]\,\text{TeV}^{-4}$, $[-0.48; 0.48]\,\text{TeV}^{-4}$, $[-0.38; 0.38]\,\text{TeV}^{-4}$ and $[-1.08; 1.08]\,\text{TeV}^{-4}$, respectively, at HL-LHC with an integrated luminosity of 3 ab$^{-1}$ \cite{Senol:2020hbh}. Comparing these phenomenological studies involving $\ell^+ \ell^- \gamma$ and $\nu\nu\gamma$ productions at the HL-LHC and our results in Eqs.~\ref{eq.22}-\ref{eq.25} at the muon collider, our sensitivity limits for anomalous couplings are approximately 54-187 times better than Ref.~\cite{Senol:2019ybv} and 15-27 times better than Ref.~\cite{Senol:2020hbh}.

These results reveal that multi-TeV muon colliders have impressive potential at future collider studies for the post-LHC and the investigation of the photon-induced process adds a new perspective to the phenomenology of muon collider.

\begin{acknowledgments}
The numerical calculations reported in this paper were partially performed at TUBITAK ULAKBIM, High Performance and Grid Computing Center (TRUBA resources).
\end{acknowledgments}

\end{document}